\documentclass[sigplan,screen]{acmart}

\usepackage[normalem]{ulem}
\usepackage{balance}
\usepackage{natbib}
\usepackage{amsmath,amsfonts,bm,mathtools,upgreek}
\usepackage{tikz}
\usepackage{comment}
\usepackage{siunitx}
\usepackage{algorithmic}
\usepackage{fontawesome5}
\usepackage{graphicx}
\usepackage{textcomp}
\usepackage{multirow}
\usepackage{xcolor}
\usepackage{xspace}
\usepackage{soul}
\usepackage{caption}
\usepackage{subcaption}
\usepackage{listings}
\usepackage{courier}
\usepackage{numprint}
\usepackage{textcomp}
\usepackage{threeparttable}
\usepackage[leftline=false,
            rightline=false,
            linewidth=.6pt,
            innerrightmargin=0pt,
            innerleftmargin=0pt,
            innertopmargin=0pt,
            innerbottommargin=0pt
]{mdframed}
\usepackage{etoolbox}
\usepackage{float}
\usepackage{caption}
\usepackage{array}

\copyrightyear{2024}
\acmYear{2024}
\setcopyright{acmlicensed}\acmConference[ASPLOS '24]{29th ACM International Conference on Architectural Support for Programming Languages and Operating Systems, Volume 2}{April 27-May 1, 2024}{La Jolla, CA, USA}
\acmBooktitle{29th ACM International Conference on Architectural Support for Programming Languages and Operating Systems, Volume 2 (ASPLOS '24), April 27-May 1, 2024, La Jolla, CA, USA}
\acmDOI{10.1145/3620665.3640403}
\acmISBN{979-8-4007-0385-0/24/04}

\definecolor{dkred}{rgb}{0.75,0,0}
\definecolor{dkgreen}{rgb}{0,0.75,0}
\definecolor{dkblue}{rgb}{0,0,0.75}
\definecolor{gray}{rgb}{0.5,0.5,0.5}
\definecolor{mauve}{rgb}{0.58,0,0.82}
\definecolor{bke}{rgb}{1,1,1}
\definecolor{darkorange}{rgb}{1.0, 0.55, 0.0}

\mdfdefinestyle{mdstyle}{
    backgroundcolor=gray!10,
    innerleftmargin=0.1cm,
    innerrightmargin=0.1cm,
    innertopmargin=0.1cm,
    innerbottommargin=0.1cm,
    leftline=true,
    rightline=true,
}

\numberwithin{equation}{section} %

\npthousandsep{,}
\newcommand{\parans}[1]{\noindent{\textbf{#1.}}} %
\newcommand{\para}[1]{\vspace{1mm}\noindent{\textbf{#1.}}} %
\newcommand{\subpara}[1]{\vspace{1mm}\noindent{\textit{#1.}}} %

\newcommand*\circled[1]{\tikz[baseline=(char.base)]{
            \node[shape=circle,draw,inner sep=.7pt] (char) {#1};}}

\DeclareSIUnit{\byte}{B}
\DeclareSIUnit{\bit}{b}
\DeclareSIUnit{\bits}{bits}
\DeclareSIUnit{\cycles}{cycles}
\DeclareSIUnit{\sets}{sets}

\newcommand{\FlushReload}{Flush+Reload\xspace}
\newcommand{\FlushFlush}{Flush+Flush\xspace}
\newcommand{\EvictReload}{Evict+Reload\xspace}
\newcommand{\PrimeProbe}{Prime+Probe\xspace}
\newcommand{\PrimeScope}{Prime+Scope\xspace}

\newcommand{\GT}{\textsc{Gt}\xspace}
\newcommand{\PS}{\textsc{Ps}\xspace}
\newcommand{\GTOpt}{\textsc{GtOp}\xspace}
\newcommand{\PSOpt}{\textsc{PsOp}\xspace}
\newcommand{\PSBest}{\textsc{PsBst}\xspace}
\newcommand{\Ours}{\textsc{BinS}\xspace}

\newcommand{\SingleSet}{\textsc{SingleSet}\xspace}
\newcommand{\PageOffset}{\textsc{PageOffset}\xspace}
\newcommand{\WholeSF}{\textsc{WholeSys}\xspace}

\newcommand{\ParaProbe}{\textsc{Parallel}\xspace}
\newcommand{\PSFlush}{\textsc{PS-Flush}\xspace}
\newcommand{\PSAlt}{\textsc{PS-Alt}\xspace}

\newcommand{\AttackStep}[1]{\textsc{Step}~#1\xspace}
\newcommand{\AttackSteps}[1]{\textsc{Steps}~#1\xspace}

\newcolumntype{M}[1]{>{\arraybackslash}m{#1}}
\newcolumntype{C}[1]{>{\centering\arraybackslash}m{#1}}

\newcommand{\CR}{Cloud Run\xspace}

\newcommand{\CentralOne}{\emph{us-central1}\xspace}

\newcommand{\DefNumPCPU}{$2$\xspace}

\newcommand{\SKX}{Skylake-SP\xspace}
\newcommand{\SKXLTwoSets}{\num{1024}\xspace}

\newcommand{\SKXLLCSetsTotal}{\num{57344}\xspace}
\newcommand{\SKXLocalLLCSetsTotal}{\num{45056}\xspace}
\newcommand{\SKXLLCSlices}{28\xspace}

\newcommand{\SKXLTwoUncertainty}{16\xspace}
\newcommand{\SKXLLCUncertainty}{896\xspace}

\newcommand{\LocalLLCSlices}{22\xspace}

\newcommand{\ICX}{Ice Lake-SP\xspace}

\newcommand{\MAdd}{\texttt{MAdd}\xspace}
\newcommand{\MDouble}{\texttt{MDouble}\xspace}
\newcommand{\MAddOne}{\texttt{MAdd1}\xspace}
\newcommand{\MDoubleOne}{\texttt{MDouble1}\xspace}

\def\Taddr{{T_{a}}}

\def\UB{\mathit{UB}}
\def\LB{\mathit{LB}}

\newcommand{\Ci}[1]{C_{#1}}

\newcommand{\testEV}{TestEviction\xspace}

\newcommand{\EVConfigGCPNumDays}{five\xspace}
\newcommand{\EVConfigGCPTotalNumExps}{$\num{1767}$\xspace}
\newcommand{\EVConfigGCPTotalNumSingleSetMeasures}{$\num{88350}$\xspace}
\newcommand{\EVConfigGCPTotalNumPageOffsetMeasures}{$\num{8835}$\xspace}
\newcommand{\EVConfigGCPTotalNumWholeSetMeasures}{$\num{1767}$\xspace}

\newcommand{\EVConfigLocalSingleEVMeasures}{$\num{1000}$\xspace}
\newcommand{\EVConfigLocalPageOffsetMeasures}{$\num{100}$\xspace}
\newcommand{\EVConfigLocalWholeSetMeasures}{$\num{20}$\xspace}
\newcommand{\EVConfigNumPageOffsets}{$\num{50}$\xspace}
\newcommand{\EVConfigNumLLCIntervalRecs}{$\num{1000}$\xspace}
\newcommand{\EVConfigVWindowsMeasures}{$\num{100}$\xspace}
\newcommand{\EVConfigVWindowsWarmups}{$\num{10}$\xspace}

\newcommand{\EVConfigMaxBacktracks}{$20$\xspace}
\newcommand{\EVConfigMaxRetries}{$10$\xspace}
\newcommand{\EVConfigSingleEVTimeout}{$\SI{100}{\milli\second}$\xspace}
\newcommand{\EVConfigSingleEVNoFilterTimeout}{$\SI{1000}{\milli\second}$\xspace}

\newcommand{\EVResGCPLLCArrival}{$11.5$\xspace}

\newcommand{\EVResLocalLLCArrival}{$0.29$\xspace}

\newcommand{\EVResLocSequentialAvgReduction}{$26.9\%$\xspace}
\newcommand{\EVResSequentialExpDuration}{$\SI{4.6}{\milli\second}$\xspace}
\newcommand{\EVResExpDurationNumAcc}{$53.0$\xspace}
\newcommand{\EVResLocParallelAvgReduction}{$42.1\%$\xspace}
\newcommand{\EVResParallelExpDuration}{$\SI{134.8}{\micro\second}$\xspace}

\newcommand{\EVResParallelProbNoAccess}{$18.4\%$\xspace}

\newcommand{\EVResGCPNoFilterPSSuccRate}{$3.2\%$\xspace}

\newcommand{\EVResGCPNoFilterPSAvgDuration}{$\SI{580}{\milli\second}$\xspace}
\newcommand{\EVResGCPNoFilterPSMedianDuration}{$\SI{504}{\milli\second}$\xspace}
\newcommand{\EVResGCPNoFilterPSStdDuration}{$\SI{329}{\milli\second}$\xspace}

\newcommand{\EVResGCPNoFilterPSOptSuccRate}{$6.9\%$\xspace}

\newcommand{\EVResGCPNoFilterPSOptAvgDuration}{$\SI{572}{\milli\second}$\xspace}
\newcommand{\EVResGCPNoFilterPSOptMedianDuration}{$\SI{495}{\milli\second}$\xspace}
\newcommand{\EVResGCPNoFilterPSOptStdDuration}{$\SI{331}{\milli\second}$\xspace}

\newcommand{\EVResGCPNoFilterVilaSuccRate}{$39.4\%$\xspace}

\newcommand{\EVResGCPNoFilterVilaAvgDuration}{$\SI{714}{\milli\second}$\xspace}

\newcommand{\EVResGCPNoFilterVilaStdDuration}{$\SI{476}{\milli\second}$\xspace}

\newcommand{\EVResGCPNoFilterVilaOptBacktracks}{$\num{32.2}$\xspace}

\newcommand{\EVResGCPNoFilterVilaOptSuccRate}{$56.0\%$\xspace}

\newcommand{\EVResGCPNoFilterVilaOptAvgDuration}{$\SI{512}{\milli\second}$\xspace}
\newcommand{\EVResGCPNoFilterVilaOptMedianDuration}{$\SI{384}{\milli\second}$\xspace}
\newcommand{\EVResGCPNoFilterVilaOptStdDuration}{$\SI{457}{\milli\second}$\xspace}

\newcommand{\EVResGCPNoFilterVilaOptDurationOffset}{$13.7$ minutes\xspace}
\newcommand{\EVResGCPNoFilterVilaOptDurationTotal}{$14.6$ hours\xspace}

\newcommand{\EVResGCPMidNightNoFilterPSSuccRate}{$3.7\%$\xspace}

\newcommand{\EVResGCPMidNightNoFilterPSAvgDuration}{$\SI{581}{\milli\second}$\xspace}
\newcommand{\EVResGCPMidNightNoFilterPSMedianDuration}{$\SI{509}{\milli\second}$\xspace}
\newcommand{\EVResGCPMidNightNoFilterPSStdDuration}{$\SI{327}{\milli\second}$\xspace}

\newcommand{\EVResGCPMidNightNoFilterPSOptSuccRate}{$7.6\%$\xspace}

\newcommand{\EVResGCPMidNightNoFilterPSOptAvgDuration}{$\SI{576}{\milli\second}$\xspace}
\newcommand{\EVResGCPMidNightNoFilterPSOptMedianDuration}{$\SI{502}{\milli\second}$\xspace}
\newcommand{\EVResGCPMidNightNoFilterPSOptStdDuration}{$\SI{332}{\milli\second}$\xspace}

\newcommand{\EVResGCPMidNightNoFilterVilaSuccRate}{$41.4\%$\xspace}

\newcommand{\EVResGCPMidNightNoFilterVilaAvgDuration}{$\SI{693}{\milli\second}$\xspace}

\newcommand{\EVResGCPMidNightNoFilterVilaStdDuration}{$\SI{482}{\milli\second}$\xspace}

\newcommand{\EVResGCPMidNightNoFilterVilaOptSuccRate}{$57.2\%$\xspace}

\newcommand{\EVResGCPMidNightNoFilterVilaOptAvgDuration}{$\SI{499}{\milli\second}$\xspace}
\newcommand{\EVResGCPMidNightNoFilterVilaOptMedianDuration}{$\SI{350}{\milli\second}$\xspace}
\newcommand{\EVResGCPMidNightNoFilterVilaOptStdDuration}{$\SI{456}{\milli\second}$\xspace}

\newcommand{\EVResLocalNoFilterPSSuccRate}{$98.5\%$\xspace}

\newcommand{\EVResLocalNoFilterPSAvgDuration}{$\SI{55.9}{\milli\second}$\xspace}
\newcommand{\EVResLocalNoFilterPSMedianDuration}{$\SI{23.8}{\milli\second}$\xspace}
\newcommand{\EVResLocalNoFilterPSStdDuration}{$\SI{166}{\milli\second}$\xspace}

\newcommand{\EVResLocalNoFilterPSOptSuccRate}{$98.2\%$\xspace}

\newcommand{\EVResLocalNoFilterPSOptAvgDuration}{$\SI{54.9}{\milli\second}$\xspace}
\newcommand{\EVResLocalNoFilterPSOptMedianDuration}{$\SI{21.7}{\milli\second}$\xspace}
\newcommand{\EVResLocalNoFilterPSOptStdDuration}{$\SI{156}{\milli\second}$\xspace}

\newcommand{\EVResLocalNoFilterVilaSuccRate}{$97.0\%$\xspace}

\newcommand{\EVResLocalNoFilterVilaAvgDuration}{$\SI{32.9}{\milli\second}$\xspace}
\newcommand{\EVResLocalNoFilterVilaMedianDuration}{$\SI{18.5}{\milli\second}$\xspace}
\newcommand{\EVResLocalNoFilterVilaStdDuration}{$\SI{72}{\milli\second}$\xspace}

\newcommand{\EVResLocalNoFilterVilaOptBacktracks}{$\num{4.0}$\xspace}

\newcommand{\EVResLocalNoFilterVilaOptSuccRate}{$98.8\%$\xspace}

\newcommand{\EVResLocalNoFilterVilaOptAvgDuration}{$\SI{21.1}{\milli\second}$\xspace}
\newcommand{\EVResLocalNoFilterVilaOptMedianDuration}{$\SI{13.7}{\milli\second}$\xspace}
\newcommand{\EVResLocalNoFilterVilaOptStdDuration}{$\SI{35}{\milli\second}$\xspace}

\newcommand{\EVResGCPFilterPSOptSuccRate}{$97.2\%$\xspace}

\newcommand{\EVResGCPFilterPSOptAvgDuration}{$\SI{33.2}{\milli\second}$\xspace}
\newcommand{\EVResGCPFilterPSOptMedianDuration}{$\SI{26.7}{\milli\second}$\xspace}
\newcommand{\EVResGCPFilterPSOptStdDuration}{$\SI{21.4}{\milli\second}$\xspace}

\newcommand{\EVResGCPFilterStrawSuccRate}{$98.1\%$\xspace}

\newcommand{\EVResGCPFilterStrawAvgDuration}{$\SI{26.6}{\milli\second}$\xspace}
\newcommand{\EVResGCPFilterStrawMedianDuration}{$\SI{23.9}{\milli\second}$\xspace}
\newcommand{\EVResGCPFilterStrawStdDuration}{$\SI{11.6}{\milli\second}$\xspace}

\newcommand{\EVResGCPFilterVilaSuccRate}{$96.7\%$\xspace}

\newcommand{\EVResGCPFilterVilaAvgDuration}{$\SI{28.8}{\milli\second}$\xspace}
\newcommand{\EVResGCPFilterVilaMedianDuration}{$\SI{25.1}{\milli\second}$\xspace}
\newcommand{\EVResGCPFilterVilaStdDuration}{$\SI{14.4}{\milli\second}$\xspace}

\newcommand{\EVResGCPFilterVilaOptSuccRate}{$97.7\%$\xspace}

\newcommand{\EVResGCPFilterVilaOptAvgDuration}{$\SI{27.2}{\milli\second}$\xspace}
\newcommand{\EVResGCPFilterVilaOptMedianDuration}{$\SI{24.7}{\milli\second}$\xspace}
\newcommand{\EVResGCPFilterVilaOptStdDuration}{$\SI{10.8}{\milli\second}$\xspace}

\newcommand{\EVResLocalFilterPSOptSuccRate}{$99.2\%$\xspace}

\newcommand{\EVResLocalFilterPSOptAvgDuration}{$\SI{14.7}{\milli\second}$\xspace}
\newcommand{\EVResLocalFilterPSOptMedianDuration}{$\SI{14.5}{\milli\second}$\xspace}
\newcommand{\EVResLocalFilterPSOptStdDuration}{$\SI{0.8}{\milli\second}$\xspace}

\newcommand{\EVResLocalFilterStrawSuccRate}{$99.9\%$\xspace}

\newcommand{\EVResLocalFilterStrawAvgDuration}{$\SI{14.1}{\milli\second}$\xspace}
\newcommand{\EVResLocalFilterStrawMedianDuration}{$\SI{13.9}{\milli\second}$\xspace}
\newcommand{\EVResLocalFilterStrawStdDuration}{$\SI{2.2}{\milli\second}$\xspace}

\newcommand{\EVResLocalFilterVilaSuccRate}{$99.3\%$\xspace}

\newcommand{\EVResLocalFilterVilaAvgDuration}{$\SI{15.2}{\milli\second}$\xspace}
\newcommand{\EVResLocalFilterVilaMedianDuration}{$\SI{14.7}{\milli\second}$\xspace}
\newcommand{\EVResLocalFilterVilaStdDuration}{$\SI{3.1}{\milli\second}$\xspace}

\newcommand{\EVResLocalFilterVilaOptSuccRate}{$99.5\%$\xspace}

\newcommand{\EVResLocalFilterVilaOptAvgDuration}{$\SI{14.7}{\milli\second}$\xspace}
\newcommand{\EVResLocalFilterVilaOptMedianDuration}{$\SI{14.4}{\milli\second}$\xspace}
\newcommand{\EVResLocalFilterVilaOptStdDuration}{$\SI{2.6}{\milli\second}$\xspace}

\newcommand{\EVResGCPPageOffsetPSAvgSFSuccRate}{$98.4\%$\xspace}

\newcommand{\EVResGCPPageOffsetPSAvgDuration}{$\SI{4.51}{\second}$\xspace}
\newcommand{\EVResGCPPageOffsetPSMedianDuration}{$\SI{3.85}{\second}$\xspace}

\newcommand{\EVResGCPPageOffsetPSStdDuration}{$\SI{2.72}{\second}$\xspace}

\newcommand{\EVResGCPPageOffsetStrawAvgSFSuccRate}{$98.0\%$\xspace}

\newcommand{\EVResGCPPageOffsetStrawAvgDuration}{$\SI{2.87}{\second}$\xspace}
\newcommand{\EVResGCPPageOffsetStrawMedianDuration}{$\SI{2.53}{\second}$\xspace}

\newcommand{\EVResGCPPageOffsetStrawStdDuration}{$\SI{1.58}{\second}$\xspace}

\newcommand{\EVResGCPPageOffsetVilaAvgSFSuccRate}{$95.6\%$\xspace}

\newcommand{\EVResGCPPageOffsetVilaAvgDuration}{$\SI{5.51}{\second}$\xspace}
\newcommand{\EVResGCPPageOffsetVilaMedianDuration}{$\SI{4.94}{\second}$\xspace}

\newcommand{\EVResGCPPageOffsetVilaStdDuration}{$\SI{2.62}{\second}$\xspace}

\newcommand{\EVResGCPPageOffsetVilaOptAvgSFSuccRate}{$97.4\%$\xspace}

\newcommand{\EVResGCPPageOffsetVilaOptAvgDuration}{$\SI{3.95}{\second}$\xspace}
\newcommand{\EVResGCPPageOffsetVilaOptMedianDuration}{$\SI{3.52}{\second}$\xspace}

\newcommand{\EVResGCPPageOffsetVilaOptStdDuration}{$\SI{1.90}{\second}$\xspace}

\newcommand{\EVResLocalPageOffsetPSOptAvgSFSuccRate}{$99.4\%$\xspace}

\newcommand{\EVResLocalPageOffsetPSOptAvgDuration}{$\SI{3.02}{\second}$\xspace}
\newcommand{\EVResLocalPageOffsetPSOptMedianDuration}{$\SI{1.39}{\second}$\xspace}

\newcommand{\EVResLocalPageOffsetPSOptStdDuration}{$\SI{2.48}{\second}$\xspace}

\newcommand{\EVResLocalPageOffsetStrawAvgSFSuccRate}{$99.5\%$\xspace}

\newcommand{\EVResLocalPageOffsetStrawAvgDuration}{$\SI{1.04}{\second}$\xspace}
\newcommand{\EVResLocalPageOffsetStrawMedianDuration}{$\SI{1.00}{\second}$\xspace}

\newcommand{\EVResLocalPageOffsetStrawStdDuration}{$\SI{0.16}{\second}$\xspace}

\newcommand{\EVResLocalPageOffsetVilaAvgSFSuccRate}{$98.6\%$\xspace}

\newcommand{\EVResLocalPageOffsetVilaAvgDuration}{$\SI{1.95}{\second}$\xspace}
\newcommand{\EVResLocalPageOffsetVilaMedianDuration}{$\SI{1.77}{\second}$\xspace}

\newcommand{\EVResLocalPageOffsetVilaStdDuration}{$\SI{0.72}{\second}$\xspace}

\newcommand{\EVResLocalPageOffsetVilaOptAvgSFSuccRate}{$99.2\%$\xspace}

\newcommand{\EVResLocalPageOffsetVilaOptAvgDuration}{$\SI{1.48}{\second}$\xspace}
\newcommand{\EVResLocalPageOffsetVilaOptMedianDuration}{$\SI{1.43}{\second}$\xspace}

\newcommand{\EVResLocalPageOffsetVilaOptStdDuration}{$\SI{0.17}{\second}$\xspace}

\newcommand{\EVResGCPWholeSetPSAvgSFSuccRate}{$91.7\%$\xspace}

\newcommand{\EVResGCPWholeSetPSMedianSFSuccRate}{$99.4\%$\xspace}

\newcommand{\EVResGCPWholeSetPSAvgDuration}{$\SI{244.4}{\second}$\xspace}
\newcommand{\EVResGCPWholeSetPSMedianDuration}{$\SI{229.6}{\second}$\xspace}

\newcommand{\EVResGCPWholeSetPSStdDuration}{$\SI{58.9}{\second}$\xspace}

\newcommand{\EVResGCPWholeSetStrawAvgSFSuccRate}{$92.6\%$\xspace}

\newcommand{\EVResGCPWholeSetStrawMedianSFSuccRate}{$99.1\%$\xspace}

\newcommand{\EVResGCPWholeSetStrawAvgDurationMin}{$2.4$ minutes\xspace}

\newcommand{\EVResGCPWholeSetStrawAvgDuration}{$\SI{142.4}{\second}$\xspace}
\newcommand{\EVResGCPWholeSetStrawMedianDuration}{$\SI{134.2}{\second}$\xspace}

\newcommand{\EVResGCPWholeSetStrawStdDuration}{$\SI{34.8}{\second}$\xspace}

\newcommand{\EVResGCPWholeSetVilaAvgSFSuccRate}{$88.1\%$\xspace}

\newcommand{\EVResGCPWholeSetVilaMedianSFSuccRate}{$96.7\%$\xspace}

\newcommand{\EVResGCPWholeSetVilaAvgDuration}{$\SI{301.1}{\second}$\xspace}
\newcommand{\EVResGCPWholeSetVilaMedianDuration}{$\SI{290.1}{\second}$\xspace}

\newcommand{\EVResGCPWholeSetVilaStdDuration}{$\SI{63.0}{\second}$\xspace}

\newcommand{\EVResGCPWholeSetVilaOptAvgSFSuccRate}{$90.5\%$\xspace}

\newcommand{\EVResGCPWholeSetVilaOptMedianSFSuccRate}{$98.5\%$\xspace}

\newcommand{\EVResGCPWholeSetVilaOptAvgDuration}{$\SI{212.6}{\second}$\xspace}
\newcommand{\EVResGCPWholeSetVilaOptMedianDuration}{$\SI{200.4}{\second}$\xspace}

\newcommand{\EVResGCPWholeSetVilaOptStdDuration}{$\SI{52.1}{\second}$\xspace}

\newcommand{\EVResLocalWholeSetPSAvgSFSuccRate}{$99.5\%$\xspace}

\newcommand{\EVResLocalWholeSetPSAvgDuration}{$\SI{175.0}{\second}$\xspace}
\newcommand{\EVResLocalWholeSetPSMedianDuration}{$\SI{185.6}{\second}$\xspace}

\newcommand{\EVResLocalWholeSetPSStdDuration}{$\SI{72.7}{\second}$\xspace}

\newcommand{\EVResLocalWholeSetStrawAvgSFSuccRate}{$99.5\%$\xspace}

\newcommand{\EVResLocalWholeSetStrawAvgDuration}{$\SI{50.1}{\second}$\xspace}
\newcommand{\EVResLocalWholeSetStrawMedianDuration}{$\SI{48.9}{\second}$\xspace}

\newcommand{\EVResLocalWholeSetStrawStdDuration}{$\SI{5.5}{\second}$\xspace}

\newcommand{\EVResLocalWholeSetVilaAvgSFSuccRate}{$99.0\%$\xspace}

\newcommand{\EVResLocalWholeSetVilaAvgDuration}{$\SI{103.6}{\second}$\xspace}
\newcommand{\EVResLocalWholeSetVilaMedianDuration}{$\SI{96.8}{\second}$\xspace}

\newcommand{\EVResLocalWholeSetVilaStdDuration}{$\SI{16.1}{\second}$\xspace}

\newcommand{\EVResLocalWholeSetVilaOptAvgSFSuccRate}{$99.1\%$\xspace}

\newcommand{\EVResLocalWholeSetVilaOptAvgDuration}{$\SI{79.6}{\second}$\xspace}
\newcommand{\EVResLocalWholeSetVilaOptMedianDuration}{$\SI{76.9}{\second}$\xspace}

\newcommand{\EVResLocalWholeSetVilaOptStdDuration}{$\SI{7.9}{\second}$\xspace}

\newcommand{\EVResGCPCandsFilter}{\SI{435}{\milli\second}\xspace}
\newcommand{\EVResGCPPageOffsetVilaDurationOverhead}{$92\%$\xspace}

\newcommand{\EVResGCPPageOffsetVilaAccOverhead}{$162\%$\xspace}

\newcommand{\EVResGCPPageOffsetVilaOptDurationOverhead}{$38\%$\xspace}

\newcommand{\EVResGCPPageOffsetVilaOptAccOverhead}{$52\%$\xspace}

\newcommand{\EVResGCPPageOffsetPSDurationOverhead}{$57\%$\xspace}

\newcommand{\EVResGCPNoFilterVilaMedianDurationNoSpace}{$\SI{1015}{\milli\second}$}
\newcommand{\EVResGCPMidNightNoFilterVilaMedianDurationNoSpace}{$\SI{1009}{\milli\second}$}

\newcommand{\OCConfigRepeats}{$1000$\xspace}

\newcommand{\OCResSKXLTwoStrawAvgDura}{\SI{1.33}{\milli\second}\xspace}

\newcommand{\OCResSKXLTwoVilaAvgDura}{\SI{2.49}{\milli\second}\xspace}

\newcommand{\OCResSKXLTwoVilaOptAvgDura}{\SI{1.90}{\milli\second}\xspace}

\newcommand{\OCResICXSFStrawAvgDura}{\SI{1.68}{\milli\second}\xspace}

\newcommand{\OCResICXSFVilaOptAvgDura}{\SI{3.07}{\milli\second}\xspace}

\newcommand{\OCResICXSFVilaAvgDura}{\SI{3.81}{\milli\second}\xspace}

\newcommand{\OCResICXLTwoStrawAvgDura}{\SI{2.28}{\milli\second}\xspace}

\newcommand{\OCResICXLTwoVilaOptAvgDura}{\SI{8.16}{\milli\second}\xspace}

\newcommand{\OCResICXLTwoVilaAvgDura}{\SI{14.48}{\milli\second}\xspace}

\newcommand{\CovertConfigGCPNumMeasure}{$\num{4070}$\xspace}
\newcommand{\CovertConfigNumEmits}{$\num{2000}$\xspace}
\newcommand{\CovertConfigNumRepeat}{$\num{10}$\xspace}
\newcommand{\CovertConfigErrBnd}{\SI{500}{\cycles}\xspace}
\newcommand{\CovertConfigErrBndns}{\SI{250}{\nano\second}\xspace}
\newcommand{\CovertConfigCtxSwitchThresh}{$\num{20000}$\xspace}

\newcommand{\CovertResParaResolution}{$\num{118}$\xspace}
\newcommand{\CovertResParaResolutionStd}{$\num{0.7}$\xspace}
\newcommand{\CovertResParaResolutionDiff}{$\num{24}$\xspace}
\newcommand{\CovertResPsResolution}{$\num{94}$\xspace}
\newcommand{\CovertResPsResolutionStd}{$\num{0.7}$\xspace}
\newcommand{\CovertResParallelPrimeLat}{$\num{1121}$\xspace}
\newcommand{\CovertResParallelPrimeLatStd}{$\num{448}$\xspace}
\newcommand{\CovertResParallelTwoKAvgDetectionRate}{$84.1\%$\xspace}

\newcommand{\CovertResParallelHundredKAvgDetectionRate}{$91.1\%$\xspace}

\newcommand{\CovertResPSFlushPrimeLat}{$\num{6024}$\xspace}
\newcommand{\CovertResPSFlushPrimeLatStd}{$\num{990}$\xspace}
\newcommand{\CovertResPSFlushTwoKAvgDetectionRate}{$15.4\%$\xspace}

\newcommand{\CovertResPSFlushHundredKAvgDetectionRate}{$82.1\%$\xspace}

\newcommand{\CovertResPSAltPrimeLat}{$\num{2777}$\xspace}
\newcommand{\CovertResPSAltPrimeLatStd}{$\num{735}$\xspace}
\newcommand{\CovertResPSAltTwoKAvgDetectionRate}{$6.0\%$\xspace}

\newcommand{\CovertResPSAltHundredKAvgDetectionRate}{$36.9\%$\xspace}

\newcommand{\MontConfigNumPosPSDTrace}{$\num{2266}$\xspace}
\newcommand{\MontConfigNumNegPSDTrace}{$\num{120103}$\xspace}
\newcommand{\MontConfigPSDFPRate}{$0.01\%$\xspace}
\newcommand{\MontConfigPSDFNRate}{$1.02\%$\xspace}

\newcommand{\MontConfigWholeSetNumMeasure}{\num{207}\xspace}
\newcommand{\MontConfigPageOffsetNumMeasure}{\num{357}\xspace}

\newcommand{\MontResWholeSetAvgDuration}{$\SI{179.7}{\second}$\xspace}
\newcommand{\MontResWholeSetStdDuration}{$\SI{177.4}{\second}$\xspace}

\newcommand{\MontResWholeSetNineFiveDuration}{$\SI{546.6}{\second}$\xspace}
\newcommand{\MontResWholeSetSuccRate}{$73.9\%$\xspace}
\newcommand{\MontResWholeSetAvgScanRateNoSpace}{762}

\newcommand{\MontResPageOffsetAvgDuration}{$\SI{6.1}{\second}$\xspace}
\newcommand{\MontResPageOffsetStdDuration}{$\SI{6.9}{\second}$\xspace}

\newcommand{\MontResPageOffsetNineFiveDuration}{$\SI{16.1}{\second}$\xspace}
\newcommand{\MontResPageOffsetSuccRate}{$94.1\%$\xspace}
\newcommand{\MontResPageOffsetAvgScanRateNoSpace}{831}

\newcommand{\MontAvgIterationDuration}{\SI{9700}{\cycles}\xspace}
\newcommand{\MontAvgHalfIterationDuration}{\SI{4850}{\cycles}\xspace}
\newcommand{\MontAvgHalfIterationDurationNoUnit}{\num{4850}}
\newcommand{\MontAvgPSDBaseFrequency}{\SI{0.41}{\mega\hertz}\xspace}
\newcommand{\MontExecutionRatio}{$25\%$\xspace}

\newcommand{\GCPSKXClockSpeed}{\SI{2}{\giga\hertz}\xspace}

\newcommand{\MontNumPairs}{$\num{52}$\xspace}
\newcommand{\MontNumPairsSignal}{$\num{47}$\xspace}
\newcommand{\MontNumTraces}{$\num{470}$\xspace}
\newcommand{\MontAvgNonceRecover}{$68\%$\xspace}
\newcommand{\MontMedianNonceRecover}{$81\%$\xspace}
\newcommand{\MontAvgBitErrorRate}{$3\%$\xspace}
\newcommand{\MontAvgEndToEndDuration}{$19$ seconds\xspace}

\newcommand{\GCPVilaOptNoFilterPageOffsetMeasures}{95\xspace}
\newcommand{\GCPVilaOptNoFilterPageOffsetAcc}{37.3\%\xspace}
\newcommand{\GCPVilaOptNoFilterPageOffsetAvgDuration}{$9.9$ minutes\xspace}

\newcommand{\GCPVilaOptNoFilterWholeSetMeasures}{69\xspace}
\newcommand{\GCPVilaOptNoFilterWholeSetBestEVs}{\num{3741}\xspace}
\newcommand{\GCPVilaOptNoFilterWholeSetAvgEVs}{\num{1074}\xspace}

\newcommand{\GCPLTwoFilterSingleSetAvgDuration}{\SI{22.3}{\milli\second}\xspace}

\newcommand{\EVResLocalFilterVilaAvgDurationNoCandsFilter}{$\SI{2.23}{\milli\second}$\xspace}
\newcommand{\EVResLocalFilterVilaOptAvgDurationNoCandsFilter}{$\SI{1.77}{\milli\second}$\xspace}
\newcommand{\EVResLocalFilterStrawAvgDurationNoCandsFilter}{$\SI{1.17}{\milli\second}$\xspace}

\def\extended{}

\ifdefined\extended
\setcopyright{none}
\renewcommand\footnotetextcopyrightpermission[1]{} %
\fi

\begin{document}

\ifdefined\extended
\title[Last-Level Cache Side-Channel Attacks Are Feasible in the Modern Public Cloud (Extended Version)]
{Last-Level Cache Side-Channel Attacks Are Feasible\\in the Modern Public Cloud (Extended Version)}
\else
\title[Last-Level Cache Side-Channel Attacks Are Feasible in the Modern Public Cloud]
{Last-Level Cache Side-Channel Attacks Are Feasible\\in the Modern Public Cloud}
\fi

\author{Zirui Neil Zhao}
\affiliation{
  \institution{University of Illinois Urbana-Champaign, USA}
}
\email{ziruiz6@illinois.edu}

\author{Adam Morrison}
\affiliation{
  \institution{Tel Aviv University, Israel}
}
\email{mad@cs.tau.ac.il}

\author{Christopher W. Fletcher}
\affiliation{
  \institution{University of Illinois Urbana-Champaign, USA}
}
\email{cwfletch@illinois.edu}

\author{Josep Torrellas}
\affiliation{
  \institution{University of Illinois Urbana-Champaign, USA}
}
\email{torrella@illinois.edu}

\renewcommand{\shortauthors}{Z. N. Zhao, A. Morrison, C. W. Fletcher, J. Torrellas}

\begin{abstract}
Last-level cache side-channel
attacks have been mostly demonstrated in highly-controlled, quiescent local environments.
Hence, it
is unclear whether such attacks are feasible in a  production cloud environment.
In the cloud,
side channels are flooded with   noise  from activities of other tenants
and, in Function-as-a-Service (FaaS) workloads, the attacker has a very limited time window to
mount the attack.

In this paper, we show that such attacks  are  feasible in practice, although they
require new techniques.
We present an end-to-end, cross-tenant attack
on a vulnerable ECDSA implementation in the public FaaS Google Cloud Run environment.
We introduce several new techniques to improve every step of the attack.
First, to speed-up the generation of eviction sets, we introduce 
{\em L2-driven candidate address filtering}  and a {\em Binary Search-based}
algorithm for address pruning. Second, to monitor victim memory accesses with high time resolution,
we introduce {\em Parallel Probing}. Finally,
we leverage {\em power spectral density} from signal processing
to easily identify the victim's target cache set in the frequency domain. 
Overall, using these mechanisms,
we extract a median value of \MontMedianNonceRecover of the secret ECDSA nonce bits
from a victim container in \MontAvgEndToEndDuration on average.

\end{abstract}



\maketitle

\ifdef{\extended}{
\pagestyle{plain}
}

\vspace{-3mm}

\section{Introduction}
\label{sec:intro}

In modern public cloud environments,
mutually-distrusting tenants share the underlying physical hardware resources.
As a result, an attacker can monitor a victim tenant's secret-dependent usage
of shared resources through various microarchitectural side channels
and exfiltrate sensitive information~\cite{primeprobe,
primeprobeL1,flush_reload,flush_flush,portsmash,smotherspectre,LLCPractical,attack_dir,
drama,tlbleed,tlbdr,binoculars,memjam,LotR,ssa,cachesniper,reload_refresh,kayaalp2016high,
prime_abort,kurth2020netcat,armageddon,xiong2020leaking,wait_a_minute,zhang2012cross,
luo2023autocat,wang2019papp}.

A particularly dangerous class of attacks
is \PrimeProbe attacks on the last-level cache (LLC)~\cite{LLCPractical,ssa,
inci2015seriously,attack_dir,armageddon,drivebyGenkin,primescope}.
This is because
these
attacks do not require the attacker to share a physical core
or memory pages with the victim program.
In such an attack,
the attacker monitors the victim program's secret-dependent accesses
to one or several LLC sets.
We refer to these LLC sets as the \emph{target LLC sets}.

Mounting LLC \PrimeProbe attacks in the modern public cloud
requires several steps~\cite{LLCPractical,ssa,inci2015seriously,getOffMyCloud,everywhereAllAtOnce},
as listed in Table~\ref{tab:attack-steps}.
First, the attacker {\em co-locates} their program with the target victim program
on the same physical machine (\AttackStep{0})~\cite{everywhereAllAtOnce,
getOffMyCloud,varadarajan2015placement,xu2015measurementCoresidence}.
Second, the attacker prepares LLC channels
by constructing LLC  eviction sets  (\AttackStep{1})~\cite{LLCPractical,ssa}.
An {\em Eviction Set} for a specific LLC set is a set of
addresses that,
once accessed, can evict any cache line mapped to that LLC set~\cite{LLCPractical,ssa}.
Using an eviction set for an LLC set $s$,
the attacker can monitor victim memory accesses to $s$. %

In practice, the attacker  does not generally know the target LLC sets.
Hence, in \AttackStep{1}, the attacker needs to build
\emph{hundreds to tens of thousands} of eviction sets,
each corresponding to a potential target LLC set~\cite{LLCPractical,ssa,oren2015spy}.
Then, the attacker scans through the potential target LLC sets
and identifies the actual target LLC sets (\AttackStep{2}).
Finally, the attacker monitors the target LLC sets
with \PrimeProbe and exfiltrates the secret (\AttackStep{3}).
Since co-location can be achieved
using techniques discussed in our prior work~\cite{everywhereAllAtOnce},
this paper focuses on \AttackSteps{1--3},
hereafter referred to as ``attack steps.''

\begin{table}[t]
    \centering
    \caption{Steps of an LLC \PrimeProbe attack in clouds.}
    \label{tab:attack-steps}
    \small
    \setlength{\tabcolsep}{3pt}
    \begin{tabular}{|C{19mm}|M{40mm}|C{17.5mm}|}
    \hline
    \textbf{Step} & \textbf{Description} & \textbf{Discussed in} \\ \hline
    \textbf{\AttackStep{0}}. Co-location & Co-locate the attacker program on the same physical host
                                as the target victim program & \cite{everywhereAllAtOnce} \\ \hline
    \textbf{\AttackStep{1}}. Prepare LLC side channels & Construct numerous eviction sets,
                                each corresponding to a potential target LLC set & Sections~\ref{sec:noise}--\ref{sec:evset} \\ \hline
    \textbf{\AttackStep{2}}. Identify  target LLC sets & Scan  LLC sets to identify those that the victim accesses in a secret-dependent manner
                                                & Sections~\ref{sec:psd}--\ref{sec:extraction} \\ \hline
    \textbf{\AttackStep{3}}. Exfiltrate information & Monitor the target LLC sets and extract information & Sections~\ref{sec:psd}--\ref{sec:extraction} \\ \hline
    \end{tabular}
    \vspace{-3mm}
\end{table}

\vspace{-2mm}
\subsection{Challenges of LLC \PrimeProbe in Clouds}

Despite the potency of LLC \PrimeProbe attacks,
executing them in a modern public cloud environment
is challenging for several reasons.
First, the modern cloud is \emph{noisy},
as the hardware is shared
by many tenants to attain high computation density~\cite{fuerst2022memory,
zhang2021faster,tirmazi2020borg}.
In particular, the LLC is flooded with noise
created by activities of other tenants.
This noise not only interferes with eviction set construction (\AttackStep{1}),
but also poses challenges to  %
identifying the target LLC sets (\AttackStep{2}) and exfiltrating information (\AttackStep{3}).

Second, the modern cloud is \emph{dynamic}.
With cloud computing paradigms like
Function-as-a-Service (FaaS)~\cite{awsLambda,cloudrun,AzureFunction},
user workloads are typically short-lived on a host
(e.g., they last only from a few minutes to tens of minutes~\cite{wang2018peeking,
hellerstein2018serverless,cloudrunContract,awsLambdaFAQ,AzureTimeout}).
As a result, the attacker has a \emph{limited time window} to
complete \emph{all} the attack steps
while co-locating with the victim.
This challenge is exacerbated by the increased number of LLC sets
in modern processors---which require preparing more eviction sets
and monitoring   more cache sets.

Third,
the lack of huge pages in some containerized environments~\cite{cloudrun}
and the wide adoption of non-inclusive LLCs
increase the effort to execute LLC \PrimeProbe attacks in clouds~\cite{attack_dir}.
Thus, while {\.{I}}nci~et~al.~\cite{inci2015seriously,inci2015seriouslyPrePrint}
conducted an LLC \PrimeProbe attack on AWS EC2 in 2015,
their techniques are incompatible with modern clouds,
as they relied on huge pages, long-running attack steps, and inclusive LLCs.

As a result of the aforementioned challenges,
cloud vendors believe that LLC \PrimeProbe attacks are not a threat ``in the wild.''
For instance,
the security design whitepaper of Amazon's Elastic Compute Cloud (EC2)~\cite{awsEC2}
explicitly rules out LLC \PrimeProbe attacks as impractical~\cite{awsNitro}.

\vspace{-3mm}
\subsection{This Paper}

This paper refutes the belief that LLC \PrimeProbe attacks
are impractical in the noisy modern public cloud.
We demonstrate an end-to-end, cross-tenant attack on cryptographic code
(a vulnerable ECDSA implementation~\cite{mont-ossl}) on \CR~\cite{cloudrun},
an FaaS platform from Google Cloud~\cite{gcp}.
Every  step of the attack requires
new techniques to address the practical
challenges posed by the cloud. %
While our demonstrated attack targets Google Cloud Run,
the techniques that we develop %
are applicable to any modern Intel server with a non-inclusive LLC.
Therefore, we believe that multi-tenant cloud products from other vendors,
such as AWS~\cite{AWS} and Azure~\cite{Azure},
may also be susceptible to our attack techniques.

This paper makes the following contributions:

\para{\circled{1} Existing \PrimeProbe approaches fail  in the cloud}
We show that \AttackSteps{1--3} of
\PrimeProbe
in Table~\ref{tab:attack-steps} are  made harder
in the \CR environment.
In particular, we show that
state-of-the-art eviction set construction algorithms,
such as group testing~\cite{EVSetTheoryPractice,CeaserS,song2019dynamically}
and \PrimeScope~\cite{primescope},
have a low chance of successfully constructing eviction sets on \CR.
Due to the noise present,
they take $10\times$ to $24\times$ as much time as
when operating in a quiescent local setting.
Since the attacker needs to construct eviction sets for
up to tens of thousands of LLC sets within a limited time window,
this low performance makes existing eviction set construction algorithms
unsuitable for the public cloud.

\para{\circled{2} Effective construction of eviction sets in the cloud}
To speed-up the generation of eviction sets in \AttackStep{1}, we introduce:
(1) a generic optimization named {\em L2-driven candidate address filtering}
that is
applicable to all
eviction set construction algorithms, and (2)
a new
{\em Binary Search-based}
eviction set construction algorithm.
By combining these two techniques,
it takes only \EVResGCPWholeSetStrawAvgDurationMin
on average to construct eviction sets for all the \SKXLLCSetsTotal LLC sets
of an Intel \SKX machine in the noisy \CR environment,
with a median success rate of \EVResGCPWholeSetStrawMedianSFSuccRate.
In contrast, utilizing the well-optimized state-of-the-art eviction set construction algorithms,
this process is expected to take at least \EVResGCPNoFilterVilaOptDurationTotal.

\para{\circled{3} Techniques for victim monitoring and target set identification}
We develop two novel techniques for \AttackSteps{2--3}.
The first one, called {\em Parallel Probing}, enables the
monitoring of the victim's memory
accesses with high time resolution and with a quick recovery from the noise created by other tenants' accesses.
The technique probes a cache set with overlapped accesses,
thus featuring a short probe latency and a simple high-performance prime pattern.

The second technique identifies the target LLC sets in a noise-resilient manner.
This technique leverages {\em power spectral density}~\cite{welch1967use} from signal processing
to detect the victim's periodic accesses to the target LLC set in the frequency domain.
It enables the attacker to identify the target LLC set in \MontResPageOffsetAvgDuration,
with an average success rate of \MontResPageOffsetSuccRate.

\para{\circled{4} End-to-end attack in production}
Using these techniques, we showcase an \emph{end-to-end}, \emph{cross-tenant} attack
on a vulnerable ECDSA implementation~\cite{mont-ossl} on \CR.
We successfully extract a median value of
\MontMedianNonceRecover of the secret ECDSA nonce bits from a victim container. %
The complete end-to-end attack,
which includes \AttackSteps{1--3} from Table~\ref{tab:attack-steps},
takes approximately \MontAvgEndToEndDuration on average after co-locating
on the victim's host.

\para{Availability}
We open sourced our implementations at \\
\url{https://github.com/zzrcxb/LLCFeasible}.

\vspace{-2mm}
\section{Background}
\label{sec:back}

\subsection{Cache-Based Side-Channel Attacks}
\label{sec:back:cache-side-channel}

Since caches are shared between processes in different security domains, they
provide an opportunity for an attacker to exfiltrate sensitive information
about a victim process by observing their cache utilization.
This constitutes a \emph{cache-based side-channel attack}.
Such attacks can be classified into \emph{reuse-based attacks} and
\emph{contention-based attacks}~\cite{liu2014randomFill}.

Reused-based attacks rely on  shared memory between  attacker and victim,
often the consequence of memory deduplication~\cite{KSM}.
In such attacks,
the attacker monitors whether shared data are brought to the cache due to victim accesses.
Notable examples of such attacks include \FlushReload~\cite{flush_reload},
\FlushFlush~\cite{flush_flush}, and \EvictReload~\cite{cacheTemplateAttack}.
However, as memory deduplication across security domains
is disabled in the cloud~\cite{remoteDedup,awsNitro},
these attacks are inapplicable.

Contention-based attacks, such as \PrimeProbe~\cite{primeprobe,primeprobeL1},
do not require shared memory between  attacker and victim.
In a \PrimeProbe attack,
the attacker monitors the victim's memory accesses to a specific cache set $s$.
The attack begins with
the attacker \emph{priming} $s$
by filling all its ways with attacker's cache lines.
Subsequently, the attacker continuously \emph{probes} these lines,
 measuring the latency of accessing them.
If the victim accesses $s$,
it evicts one of the attacker's cache lines, which the attacker can
detect through increased probe latency.
The attacker then re-primes $s$ and repeats the probing process to continue monitoring.

Cloud vendors generally prevent processes of different tenants
from sharing the same physical core at the same time~\cite{core_scheduling,awsNitro}.
Therefore, the attacker has to perform a \emph{cross-core} attack targeting the shared LLC.
On modern processors, the LLC is split into multiple slices.
Each physical address is hashed to one of the slices.

\vspace{-2mm}
\subsection{Eviction Sets}
\label{sec:back:evset}

A key step in \PrimeProbe is the construction of
an \emph{eviction set}~\cite{LLCPractical,EVSetTheoryPractice}.
An eviction set
for a specific cache set $s$
is a set of addresses that,
once accessed, evict any cache line mapped to $s$~\cite{LLCPractical,EVSetTheoryPractice}.
Given a $W$-way cache,
an eviction set needs to contain at least $W$ addresses that are mapped to $s$.
These addresses are referred to as
\emph{congruent addresses}~\cite{EVSetTheoryPractice}.
An eviction set is \emph{minimal} if it has only $W$ congruent addresses.

\subsubsection{Eviction Set Construction Algorithms}
\label{sec:back:evset:algo}

Building a minimal eviction set for a cache set $s$
generally consists of two steps~\cite{LLCPractical,EVSetTheoryPractice}.
The first step is to create a \emph{candidate set}
that contains a sufficiently large set of \emph{candidate addresses},
of which at least $W$ addresses are congruent in $s$.
The second step is to prune the candidate set into a minimal set.

\begin{figure}[t]
    \centering
    \includegraphics[width=.9\linewidth]{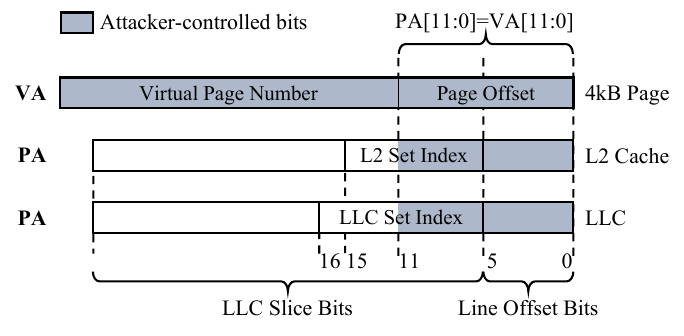}
    \vspace{-1mm}
    \caption{Mapping addresses to \SKX's L2 and LLC.
    }
    \Description{Mapping addresses to \SKX's L2 and LLC.}
    \label{fig:skx_pa_mapping}
    \vspace{-5mm}
\end{figure}

\para{\circled{1} Candidate set construction}
When a program accesses a virtual address (VA), the address
 is translated to a physical address (PA) during the access.
Part of the PA is used to determine to which cache set the PA maps.
For example,
Figure~\ref{fig:skx_pa_mapping} illustrates the address mapping of Intel \SKX's L2 and LLC.
The L2 uses PA bits 15--6 as the set index bits to map a PA to an L2 set.
The LLC uses PA bits 16--6 as the set index bits.
All the PA bits except for the low-order 6 bits are used
 to map a PA to an LLC slice~\cite{mccalpin2021LLCHash}.
The low-order 6 bits of the PA and VA are shared and are the line offset bits.
The low-order 12 bits of the PA and VA are also shared and are the page offset bits. This
is because the standard page size is \SI{4}{\kilo\byte}.

An unprivileged attacker can control only the page offset of a PA. They
lack  control and knowledge of the higher-order PA bits.
As a result,
the attacker has only partial control and knowledge of
the set index bits of the L2 and LLC,
as well as of the slice index bits of the LLC.
Therefore, for a given attacker-controlled VA,
there are a number of possible L2 or LLC sets to which it may map.
We refer to this number as the \emph{cache uncertainty}, denoted by $U$.

In general,
the set index bits are directly used as the set number.
Therefore, the L2 cache's uncertainty is $U_{L2}=2^{n_{uc}}$,
where $n_{uc}$ is the number of uncontrollable L2 set index bits.
For the sliced LLC, its uncertainty depends on the slice hash function.
On modern processors,
LLC slice bits usually map to individual slices
via a complex, non-linear hash function~\cite{attack_dir,mccalpin2021LLCHash}.
As a result, partial control over the slice index bits is not enough
to reduce the number of possible slices that a VA might hash to.
Hence, the LLC's uncertainty is $U_{LLC}=2^{n_{uc}}\times n_{slices}$,
where $n_{uc}$ is the number of uncontrollable LLC set index bits and
$n_{slices}$ is the number of LLC slices.
In the \SKX's address mapping shown in Figure~\ref{fig:skx_pa_mapping},
there are $5$ uncontrollable LLC set index bits
and $4$ uncontrollable L2 set index bits.
Hence, a 28-slice \SKX has an LLC uncertainty of
$U_{LLC}=2^{5}\times 28=896$ sets
and an L2 uncertainty of $U_{L2}=2^{4}=16$ sets.

When constructing a candidate set for a target cache set $s$ at page offset $o$,
the set needs to contain a large number of addresses with page offset $o$
due to cache uncertainty.
Intuitively, the cache uncertainty $U$ indicates how \emph{unlikely}
a candidate address  maps to $s$.
Therefore,
the greater the value of $U$,
the larger the candidate set needs to be~\cite{EVSetTheoryPractice,song2019dynamically}.

\para{\circled{2} Pruning the candidate set into a minimal eviction set}
Given a candidate set, there are several  algorithms~\cite{LLCPractical,
EVSetTheoryPractice,CeaserS,primescope,RandomCacheAnalysis,xue2023ctpp}
to build a minimal eviction set with $W$ congruent addresses.
We briefly describe two state-of-the-art
algorithms~\cite{EVSetTheoryPractice,CeaserS,primescope}.
To simplify the discussion,
we assume that there is an address $\Taddr$ that is mapped to cache set $s$
and accessible by the attacker.
Consequently, the attacker can determine if a set of addresses forms an eviction set for $s$
by testing whether they  evict $\Taddr$ after being accessed.

\vspace{1mm}
\noindent{\textit{Algorithm 1: Group testing}~\cite{EVSetTheoryPractice,CeaserS}.}
Group testing splits the candidate set into $G$ groups
of approximately the same size.
A common choice of $G$ is $G=W+1$~\cite{EVSetTheoryPractice,CeaserS}.
After the split,
the algorithm
withholds one group from the candidate set
and tests whether the remaining addresses can still evict $\Taddr$.
This process involves first loading $\Taddr$ into the cache,
traversing the remaining addresses,
and timing an access to $\Taddr$ to check if it remains cached.
If $\Taddr$ is evicted,
the withheld group is discarded and the candidate set is reduced;
otherwise, the withheld group is added back to the candidate set and
the algorithm  withholds a different group.
Overall,
with $G=W+1$,
group testing has a complexity of $O(W^2 N)$ memory accesses~\cite{EVSetTheoryPractice},
where $W$ is the associativity of the target cache and $N$ is the candidate set size.

\vspace{1mm}
\noindent{\textit{Algorithm 2: \PrimeScope}~\cite{primescope}.}
\PrimeScope first loads $\Taddr$. %
Then, it \emph{sequentially} accesses each candidate address from the list.
After each candidate address is accessed,
the algorithm checks whether $\Taddr$ is still cached. %
If it is not, that indicates that the last accessed candidate address is congruent,
and it is added to the eviction set.
This search for congruent addresses is repeated
until $W$ different congruent addresses are identified,
which form a minimal eviction set for $s$.

\vspace{-2mm}
\subsubsection{Number of Eviction Sets}
\label{sec:back:evset:num}

In practice,
a victim often accesses only a few target cache sets in a secret-dependent manner.
An unprivileged attacker, however, generally has limited or
no information about the locations of these target cache sets.
Consequently, in \AttackStep{1} of Table~\ref{tab:attack-steps},
the attacker needs to build  eviction sets
for all possible cache sets that might be the targets.
Subsequently, in \AttackStep{2},
the attacker uses \PrimeProbe to monitor each of these possible cache sets to identify
the actual target cache sets.

The number of eviction sets that the attacker needs to build and monitor
depends on how much information about the target cache sets the attacker has.
If the attacker knows the page offset of a target cache set,
they only need to build eviction sets for cache sets
corresponding to that page offset and monitor such sets~\cite{LLCPractical,oren2015spy}.
We refer to this scenario as \PageOffset.
Conversely, if the attacker has no information about the target sets,
they must construct eviction sets for
all cache sets in the system and monitor them~\cite{LLCPractical,oren2015spy}.
We refer to this scenario as \WholeSF.
Considering the standard \SI{4}{\kilo\byte} page size and
\SI{64}{\byte} cache line size,
the attacker in the \WholeSF scenario
needs to build and monitor $64\times$ as many eviction sets as in the \PageOffset scenario.
For a 28-slice \SKX CPU,
the attacker needs to build $U_{LLC}=\SKXLLCUncertainty$ eviction sets for the LLC sets at a give page offset
and $U_{LLC}\times 64=\SKXLLCSetsTotal$ eviction sets for all LLC sets in the system.

\vspace{-1mm}
\subsubsection{Bulk Eviction Set Construction}
\label{sec:back:evset:multi-ev}

The process of constructing eviction sets for the \PageOffset or \WholeSF scenarios
is based on the procedure  to build  a single eviction set.
Because one can construct eviction sets for the \WholeSF scenario
by repeating the process for the \PageOffset scenario at each possible page offset,
we only explain the generation of eviction sets for the \PageOffset scenario.

First, we build a candidate set containing addresses with the target page offset.
The candidate set needs to contain enough congruent addresses
for \emph{any cache set} at that page offset.
Second, we pick and remove one address from the candidate set
and use it as the target address $\Taddr$.
Third, we use either of the address pruning algorithms in Section~\ref{sec:back:evset:algo}
to build an eviction set for the cache set to which $\Taddr$ maps.
The constructed eviction set is removed from the candidate set and
saved to a list $L$
containing all the eviction sets that have been built so far.
Fourth, we pick and remove another address $A$ from the reduced candidate set.
If $A$ cannot be evicted by any eviction set in $L$,
we use $A$ as the target address $\Taddr$ and proceed to the third step
to construct a new eviction set;
otherwise, we discard $A$ and repeat the fourth step.
We stop when either we run out of candidate addresses or
enough eviction sets have been built.

\vspace{-1mm}
\subsection{Non-Inclusive LLC in Intel Server CPUs}
\label{sec:back:non-inc}
Beginning with the \SKX microarchitecture~\cite{skylakeSP},
Intel adopted a non-inclusive LLC design on their server platforms.
Under this design,
cache lines in private caches may or may not exist in the LLC.
The Snoop Filter (SF)~\cite{skylakeSP}
tracks the ownership of cache lines present only in private caches,
serving as a coherence directory for such cache lines.
Similar to the LLC, the SF is shared among cores and sliced.
The SF has the same  number of sets,  number of slices,
and  slice hash function as the LLC.
Therefore, if two addresses  map  to the same LLC set,
they   also map  to the same SF set.

The interactions among private caches, SF, and LLC are complex and undocumented.
Based on prior work~\cite{attack_dir} and our reverse engineering, we provide a
brief overview of these interactions, acknowledging that our descriptions may not
be entirely accurate or exhaustive.

Lines that are in state \textsc{Exclusive (E)} or \textsc{Modified (M)} in one of the
private caches are tracked by the SF; we call these lines {\em private}.
Lines that are in state \textsc{Shared (S)}
in at least one of the private caches
are tracked by the LLC (and, therefore, are also cached in the LLC);
we call these lines {\em shared}.

When an SF entry is evicted, the corresponding line in the private cache is also  evicted.
The evicted line may
be inserted into the LLC depending on a reuse predictor~\cite{skylakePMon,SPECFlags}.
When a line cached in the LLC needs to transition to state \textsc{E} or \textsc{M} due to an access,
it is removed from the LLC
and an SF entry is allocated to track it. When a private line transitions to state \textsc{S},
it is inserted into the LLC, and its  SF entry is freed.

\vspace{-2mm}
\subsection{Function-as-a-Service (FaaS) Platform}
\label{sec:back:FaaS}

Function-as-a-Service (FaaS)~\cite{cloudrun,awsLambda,AzureFunction}
is a popular cloud computing paradigm. The basic unit of execution is a
function, which executes in an ephemeral, stateless container or
micro virtual machine  created and scheduled on demand in an event-driven manner.
Applications are then composed of multiple  functions
that communicate with each other. Users upload  their functions
and the cloud provider supplies all the libraries, runtime
environment, and system services needed to run them. The FaaS platform orchestrator automatically adjusts
the number of container instances to match the function invocation demand.
These instances often have a short lifetime~\cite{cloudrunContract, awsLambdaFAQ,AzureTimeout}.
This design allows the concurrent execution of many instances
on a single physical host, improving  hardware utilization.

In this paper,
we use the FaaS Google Cloud Run platform~\cite{cloudrun}.
In our experiments, we find that
the CPU microarchitecture used in \CR datacenters is dominated by Intel \SKX and Cascade Lake-SP.
Since these two microarchitectures have similar  cache hierarchies,
we focus our discussion on \SKX.
Table~\ref{tab:skx_cache} lists the key parameters of \SKX's cache hierarchy.

\begin{table}[t]
    \caption{Parameters of the \SKX cache hierarchy.}
    \label{tab:skx_cache}
    \small
    \setlength{\tabcolsep}{3.5pt}
    \begin{tabular}{|M{12.8mm}|M{65mm}|}
    \hline
    \textbf{Structure} & \textbf{Parameters} \\ \hline
    L1           & Data/Instruction: \SI{32}{\kilo\byte}, 8 ways, 64 sets, 64 B line \\ \hline
    L2           & \SI{1}{\mega\byte}, 16 ways,  \num{1024}  sets, non-inclusive to L1 \\ \hline
    LLC Slice  & \SI{1.375}{\mega\byte}, 11 ways,  \num{2048}  sets, non-inclusive to L1/L2  \\ \hline
    SF Slice   & 12 ways, \num{2048} sets  \\ \hline
    Num. Slices & Up to 28 slices. A 28-slice LLC and SF is the most common configuration in \CR datacenters\\ \hline
    \end{tabular}
    \vspace{-5mm}
\end{table}

\vspace{-1mm}
\section{Threat Model}
\label{sec:threat}

In this paper,
we assume an attacker who aims to exfiltrate sensitive information
from a victim containerized service running
on a public FaaS platform  such as \CR~\cite{cloudrun}
through LLC side channels.
In our prior work~\cite{everywhereAllAtOnce},
we have demonstrated how an  attacker can co-locate
with a target victim container on \CR.
If the victim runs  container instances on multiple hosts,
our techniques can co-locate attacker containers with
a large portion of the victim instances.
Therefore, we assume the co-location step is completed and
 focus on
\AttackSteps{1--3} from Table~\ref{tab:attack-steps}.

We assume that the attacker is an unprivileged user of a FaaS platform.
The attacker's interaction with the FaaS platform is limited to
the standard interfaces that are available to all platform users.
Using these interfaces,
the attacker can deploy services %
that contain attacker-controlled binaries.
Finally, we assume that the attacker can trigger the victim's execution
by sending requests to the victim service,
either directly or through interaction with a public web application
that the victim service is part of.

Since cloud vendors typically prevent different users
from simultaneously using the same physical core via Simultaneous Multithreading (SMT)~\cite{core_scheduling,awsNitro},
the attacker must perform a \emph{cross-core} cache attack.
Similar to prior work~\cite{attack_dir,primescope} that targets Intel \SKX,
we create eviction sets for the SF and monitor the SF for the victim's accesses.
Note that an SF eviction set is also an LLC eviction set,
as the SF and LLC share the same set mapping
and the SF has more ways.

Lastly, we found that user containers on \CR are unable to allocate huge pages.
Therefore, we assume that the attacker can only rely on
the standard \SI{4}{\kilo\byte} pages to construct eviction sets.
This assumption is consistent with other restricted execution environments~\cite{EVSetTheoryPractice,
oren2015spy,rowhammerJS,schwarz2018javascript} and
 places fewer requirements on the attacker's capability.

\vspace{-1mm}
\section{Existing Eviction Sets Fail in the Cloud}
\label{sec:noise}

In this section, we show that existing algorithms to construct eviction sets fail
in the cloud. This is because of the noise in the environment and the reduced time
window available to construct the eviction sets. In the following, we
first examine the resilience to environmental noise of %
a core primitive used by all address pruning algorithms (Section~\ref{sec:noise:testev}).
Then, we evaluate the success rate and execution time of
the two state-of-the-art address pruning algorithms on \CR (Section~\ref{sec:noise:ev-fails}),
and investigate the reasons why they fall short in the cloud (Section~\ref{sec:noise:understand}).

\vspace{-2mm}
\subsection{\testEV Primitive \& Its Noise Susceptibility}
\label{sec:noise:testev}

All the address pruning algorithms
require a primitive that tests whether
a target cache line is
evicted from the target cache after a set of
candidate addresses are accessed~\cite{LLCPractical,EVSetTheoryPractice,CeaserS,attack_dir}.
We refer to this generic primitive as $\testEV$.
Specifically,
group testing uses $\testEV$ to prune away non-congruent addresses,
while \PrimeScope employs it to identify congruent addresses.

Due to
environmental noise,
$\testEV$ can return \emph{false-positive} results---i.e., the target cache line
is evicted by accesses from other tenants and
not by the accesses to the candidate addresses.
When this occurs in group testing,
the algorithm may discard a group of addresses with congruent addresses,
falsely believing that the remaining addresses contain enough congruent addresses.
Similarly, \PrimeScope can misidentify a non-congruent address
as a congruent one, incorrectly including it in the eviction set.
In both cases, the algorithms may fail to construct an eviction set.

In general, the longer the execution time of $\testEV$ is,
the more susceptible it becomes to noise,
due to the increased likelihood of the target cache set being accessed by
other tenants during its execution.
Thus, the execution time of $\testEV$
not only affects the end-to-end execution time of the algorithm,
but also the algorithm's resilience to noise.

Prior work~\cite{EVSetTheoryPractice} that uses the group testing algorithm
implements $\testEV$ with linked-list traversal~\cite{VilaGithub}.
As this implementation serializes memory accesses to candidate addresses,
we refer to this type of implementation as \emph{sequential \testEV}.
\PrimeScope also uses  sequential $\testEV$, %
as it tests whether a target line is still cached
after \emph{each} access to a candidate address.
Since sequential $\testEV$ does not exploit
memory-level parallelism (MLP),
it has a long execution time.

In our work,
we find that overlapping accesses to candidate addresses
to exploit MLP can significantly reduce the execution time of $\testEV$.
We refer to this implementation as \emph{parallel $\testEV$}.
It is based on a pattern proposed by Gruss~et~al.~\cite{rowhammerJS},
and our implementation can be found online~\cite{parallelTestEV}.
However,
as will be shown in Sections~\ref{sec:noise:ev-fails}~and~\ref{sec:noise:understand},
even though parallel $\testEV$ is significantly faster than sequential $\testEV$,
it alone is not enough to overcome the noise in the cloud.
In the rest of this paper,
we use parallel $\testEV$
in all algorithms
except for \PrimeScope,
which is incompatible with parallel $\testEV$ due to its algorithm design.

\vspace{-2mm}
\subsection{Noise Resilience of Existing Algorithms}
\label{sec:noise:ev-fails}

In this section, we implement both the group testing
and \PrimeScope algorithms for \SKX's SF.
We then evaluate their success rates and execution times
in a local environment with minimal noise,
as well as in the %
\CR environment,
which features a significant level of environmental noise from other tenants.

\parans{Implementation}
Following   prior work that builds SF eviction sets~\cite{attack_dir,primescope},
we first construct a minimal LLC eviction set comprising $11$ congruent
addresses,
and then expand it to an SF eviction set by finding one additional congruent address.
To insert cache lines into the LLC,
we use a helper thread running on a different physical core
that repeats the accesses made by the main thread.
These repeated accesses turn the state of the cache lines to {\em S},
and thus cause the lines to be stored in the LLC (Section~\ref{sec:back:non-inc}).
Similar techniques are used in prior work~\cite{attack_dir,primescope}.
Finally, as per Section~\ref{sec:noise:testev},
our group testing implementation uses parallel $\testEV$,
while our \PrimeScope implementation uses sequential $\testEV$.

To ensure a fair comparison among algorithms,
we re-implement group testing and \PrimeScope using the same data structures
to  store candidate sets and eviction sets,
and the same primitives to test whether a set of addresses
is an eviction set. We call these algorithms \GT and \PS, respectively.
In addition,
we also implement optimized versions of these algorithms for \SKX.
Details of these optimizations are presented
\ifdefined\extended
in Appendix~\ref{app:algo-opt}.
\else
in the extended version of this paper~\cite{LLCFeasibleExt}.
\fi
We call these optimized algorithms \GTOpt and \PSOpt.

\para{Experiment setup}
We evaluate these algorithms in both a cloud setup and a local setup.

\subpara{Cloud setup}
We deploy our attacker service to the \CentralOne data center,
where we observe the largest \CR cluster.
Since our setup requires a concurrently running helper thread,
each attacker instance requests \DefNumPCPU physical cores.
In \CentralOne, the predominant CPU model used by \CR is
the Intel Xeon Platinum 8173M, which is
a \SKX processor with \SKXLLCSlices LLC/SF slices.

During each experiment,
we launch $300$ attacker instances and
retain only one per host.
We then use each algorithm to build SF eviction sets for
\EVConfigNumPageOffsets random cache sets.
To measure the effects of environmental noise fluctuations
due to computation demand changes,
we repeat our experiments for \EVConfigGCPNumDays days
and at four different periods each day, namely,
morning (9--11am), afternoon (3--5pm),
evening (8--10pm), and early morning (3--5am).
Altogether, we conducted \EVConfigGCPTotalNumExps experiments on \CR,
totaling \EVConfigGCPTotalNumSingleSetMeasures eviction set constructions
for each algorithm.

\subpara{Local setup}
Our local setup uses a \SKX processor with the
Intel Xeon Gold 6152, which has $\LocalLLCSlices$ LLC/SF slices.
During the experiment,
the system operates with minimal activity
beyond the running attacker container instance.
We employ each algorithm to construct \EVConfigLocalSingleEVMeasures SF eviction sets.

\subpara{Algorithms}
For each SF eviction set, we allow each algorithm to make at most \EVConfigMaxRetries
construction attempts. If the algorithm fails  these many times or
it takes more than \EVConfigSingleEVNoFilterTimeout to complete, we declare its failure.
For group testing, which uses backtracking to recover
from erroneous $\testEV$ results,
we permit at most \EVConfigMaxBacktracks backtracks per attempt.

In our experiments, we need to start by generating a set of
candidate addresses for a given page offset. Empirically, we find that
a set with $3UW$ candidate addresses is enough for \SKX's LLC/SF,
where $U$ and $W$ are the cache uncertainty (Section~\ref{sec:back:evset:algo}) and associativity, respectively.

\para{Results}
Table~\ref{tab:eff_nofilter} shows the effectiveness of
the state-of-the-art algorithms to construct an eviction set for SF in different environments: quiescent local,
\CR, and \CR from 3am to 5am,
which are typically considered ``quiet hours''.
The metrics shown are the success rate, average execution time, standard deviation of execution time, and median execution time.
The success rate is
the probability of successfully constructing an SF eviction set.
The execution time measures the real-world time that it takes to reduce a candidate set
to an LLC eviction set and then extend it with one additional congruent address
to form an SF eviction set.

\begin{table}[t]
    \centering
    \caption{Effectiveness of the state-of-the-art address pruning algorithms in different environments. The metrics shown are: success rate, average execution time, standard deviation of execution time, and median execution time.
    }
    \label{tab:eff_nofilter}
    \small
    \setlength{\tabcolsep}{3pt}
    \begin{threeparttable}
        \begin{tabular}{|c|c|c|c|c|c|}
            \hline
            \rule{0pt}{2.2ex}
            \textbf{Env.} & \textbf{Metrics} & \textbf{\GT} & \textbf{\GTOpt} & \textbf{\PS} & \textbf{\PSOpt} \\ \hline
            \multirow{4}{*}{\begin{tabular}[c]{@{}c@{}}Quiescent \\ Local\end{tabular}} &
            \rule{0pt}{2.2ex} Succ. Rate
            & \EVResLocalNoFilterVilaSuccRate &
            \EVResLocalNoFilterVilaOptSuccRate & \EVResLocalNoFilterPSSuccRate & \EVResLocalNoFilterPSOptSuccRate \\ \cline{2-6}
                & \rule{0pt}{2.2ex} Avg. Time    & \EVResLocalNoFilterVilaAvgDuration &
            \EVResLocalNoFilterVilaOptAvgDuration  & \EVResLocalNoFilterPSAvgDuration & \EVResLocalNoFilterPSOptAvgDuration \\ \cline{2-6}
                & \rule{0pt}{2.2ex} Stddev Time    & \EVResLocalNoFilterVilaStdDuration &
                \EVResLocalNoFilterVilaOptStdDuration  & \EVResLocalNoFilterPSStdDuration & \EVResLocalNoFilterPSOptStdDuration \\ \cline{2-6}
                & \rule{0pt}{2.2ex} Med. Time    & \EVResLocalNoFilterVilaMedianDuration &
            \EVResLocalNoFilterVilaOptMedianDuration & \EVResLocalNoFilterPSMedianDuration & \EVResLocalNoFilterPSOptMedianDuration \\ \hline\hline
            \multirow{4}{*}{\begin{tabular}[c]{@{}c@{}}Cloud\\ Run\end{tabular}} &
            \rule{0pt}{2.2ex} Succ. Rate & \EVResGCPNoFilterVilaSuccRate &
            \EVResGCPNoFilterVilaOptSuccRate & \EVResGCPNoFilterPSSuccRate & \EVResGCPNoFilterPSOptSuccRate \\ \cline{2-6}
                & \rule{0pt}{2.2ex} Avg. Time   & \EVResGCPNoFilterVilaAvgDuration &
            \EVResGCPNoFilterVilaOptAvgDuration  & \EVResGCPNoFilterPSAvgDuration & \EVResGCPNoFilterPSOptAvgDuration \\ \cline{2-6}
                & \rule{0pt}{2.2ex} Stddev Time   & \EVResGCPNoFilterVilaStdDuration &
                \EVResGCPNoFilterVilaOptStdDuration  & \EVResGCPNoFilterPSStdDuration & \EVResGCPNoFilterPSOptStdDuration \\ \cline{2-6}
                & \rule{0pt}{2.2ex} Med. Time   & \EVResGCPNoFilterVilaMedianDurationNoSpace &
            \EVResGCPNoFilterVilaOptMedianDuration & \EVResGCPNoFilterPSMedianDuration & \EVResGCPNoFilterPSOptMedianDuration \\ \hline
            \hline\multirow{4}{*}{\begin{tabular}[c]{@{}c@{}}Cloud\\ Run \\ (3-5am)\end{tabular}} &
            \rule{0pt}{2.2ex} Succ. Rate & \EVResGCPMidNightNoFilterVilaSuccRate &
            \EVResGCPMidNightNoFilterVilaOptSuccRate & \EVResGCPMidNightNoFilterPSSuccRate & \EVResGCPMidNightNoFilterPSOptSuccRate \\ \cline{2-6}
                & \rule{0pt}{2.2ex} Avg. Time   & \EVResGCPMidNightNoFilterVilaAvgDuration &
            \EVResGCPMidNightNoFilterVilaOptAvgDuration  & \EVResGCPMidNightNoFilterPSAvgDuration & \EVResGCPMidNightNoFilterPSOptAvgDuration \\ \cline{2-6}
                & \rule{0pt}{2.2ex} Stddev Time   & \EVResGCPMidNightNoFilterVilaStdDuration &
            \EVResGCPMidNightNoFilterVilaOptStdDuration  & \EVResGCPMidNightNoFilterPSStdDuration & \EVResGCPMidNightNoFilterPSOptStdDuration \\ \cline{2-6}
                & \rule{0pt}{2.2ex} Med. Time   & \EVResGCPMidNightNoFilterVilaMedianDurationNoSpace &
            \EVResGCPMidNightNoFilterVilaOptMedianDuration & \EVResGCPMidNightNoFilterPSMedianDuration & \EVResGCPMidNightNoFilterPSOptMedianDuration \\ \hline
        \end{tabular}
    \end{threeparttable}
    \vspace{-3mm}
\end{table}

We see that all algorithms achieve very high success rates and
good performance in the quiescent local environment.
However, on \CR, where there is substantial environmental noise from other tenants,
all algorithms suffer significant degradation in both success rate and performance.
Moreover, we do not observe significant variations in success rate or execution time
across different periods of a day, including the 3am to 5am quiet hours.
We believe this could be due to certain server consolidation mechanisms
that adjust the number of active hosts based on demand~\cite{thresholdConsolidation,
EnaCloud,EnergyManagementInDC},
leading to a relatively constant load level on  active hosts throughout the day.

\para{Implications}
As discussed in Section~\ref{sec:back:evset:num},
an unprivileged attacker needs to construct eviction sets for
all SF sets at a given page offset (\PageOffset) or in the system (\WholeSF).
We estimate the time to construct  many eviction sets
as $n_{sets}\times t_{avg}/SR$,
where $n_{sets}$ is the number of eviction sets we need to build,
$t_{avg}$ is the average execution time of attempting to construct one eviction set,
and $SR$ is the success rate.
Similar metrics are also used in prior work~\cite{song2019dynamically}.

For the \SKX processor that we are targeting,
the attacker needs to build \SKXLLCUncertainty and \SKXLLCSetsTotal SF eviction sets
in the \PageOffset and \WholeSF scenarios respectively (Section~\ref{sec:back:evset:num}).
Hence, on \CR, \GTOpt, the fastest and  most noise-resilient of the evaluated algorithms,
would take \EVResGCPNoFilterVilaOptDurationOffset and \EVResGCPNoFilterVilaOptDurationTotal
to construct eviction sets required in the \PageOffset and \WholeSF scenarios, respectively.

We performed two additional small-scale experiments to validate our estimation.
In the first experiment,
which is conducted on \GCPVilaOptNoFilterPageOffsetMeasures hosts,
\GTOpt attempts to construct the \SKXLLCUncertainty eviction sets
required in the \PageOffset scenario. \GTOpt takes, on
average, \GCPVilaOptNoFilterPageOffsetAvgDuration to complete the task, and it
only succeeds in \GCPVilaOptNoFilterPageOffsetAcc
of the sets.
In the second experiment,
which is conducted on \GCPVilaOptNoFilterWholeSetMeasures hosts,
\GTOpt tries to construct the \SKXLLCSetsTotal eviction sets
required in the \WholeSF scenario.
Due to the timeout constraint of \CR~\cite{cloudrunTimeout},
we can only run \GTOpt for one hour and
thus report the number of eviction sets it constructs under the constraint.
Our best outcome is constructing \GCPVilaOptNoFilterWholeSetBestEVs
eviction sets in one hour,
with an average number of \GCPVilaOptNoFilterWholeSetAvgEVs sets in one hour.
This means that building eviction sets for the system's \SKXLLCSetsTotal
SF sets would take \GTOpt over
$\SKXLLCSetsTotal/\GCPVilaOptNoFilterWholeSetBestEVs\approx15$
hours even in the best case.

This performance is unsatisfactory for a practical attack on FaaS platforms
for several reasons.
First, on some popular FaaS platforms~\cite{AzureFunction,awsLambda},
the attacker can only execute for $10$ to $15$ minutes
before timeout~\cite{awsLambdaFAQ,AzureTimeout}.
Even on a more permissive platform like \CR,
the maximum timeout is just one hour~\cite{cloudrunTimeout}.
After a timeout, the attacker might \emph{not} reconnect to the same instance~\cite{cloudrunTimeout},
thus losing the attack progress.
Second, container instances usually have a short lifetime
before being terminated~\cite{wang2018peeking,hellerstein2018serverless}.
Hence, the long eviction set construction time means that the co-located
victim instance may get terminated before eviction sets are ready.
Finally, as FaaS platforms charge customers by the CPU time,
the long execution time can cause significant financial cost
to the attacker.
This is especially the case if the attacker is launching many attacker instances
on different hosts to increase the chance of a successful attack.

\vspace{-2mm}
\subsection{Explaining the Results}
\label{sec:noise:understand}

Compared to a quiescent local environment,
we find that the cloud environment has a drastically higher rate of LLC accesses
made by other tenants, and that $\testEV$'s execution is slower.
These two factors contribute to why the state-of-the-art algorithms are ineffective
in a cloud environment.
Our conclusion is based on the following two experiments.
To ensure meaningful comparisons,
both experiments are conducted using the container instances of Section~\ref{sec:noise:ev-fails}.

\para{Experiment 1: LLC set access frequency}
In this experiment,
we measure how frequently an LLC set is accessed by background activities,
such as system processes and processes of other tenants.
The reason why we focus on the access frequency of the LLC
instead of the SF is because address pruning algorithms
build eviction sets in the LLC and then expand them to SF eviction sets
(Section~\ref{sec:noise:ev-fails}).

During the experiment,
we first construct an eviction set for a randomly chosen LLC set.
Then, we detect background LLC accesses with \PrimeProbe~\cite{primeprobe}.
We record the timestamp of each  LLC access.
Each experiment trial collects the timestamps of
$\EVConfigNumLLCIntervalRecs$ back-to-back LLC accesses.
On \CR, we perform $\EVConfigNumPageOffsets$ trials per host
($\EVConfigGCPTotalNumSingleSetMeasures$ trials in total).
In the local environment,
we carry out $\EVConfigLocalSingleEVMeasures$ trials.

\para{Experiment 2: \testEV execution duration}
In this experiment,
we measure the execution duration of both the parallel and sequential $\testEV$
when testing varying numbers of candidate addresses.
We perform the measurement in both the \CR and local environments.
For each host and candidate set size,
we measure $\testEV$'s execution time for
\EVConfigVWindowsMeasures times after \EVConfigVWindowsWarmups warm-ups.

\begin{figure}[t]
    \centering
    \includegraphics[width=\linewidth]{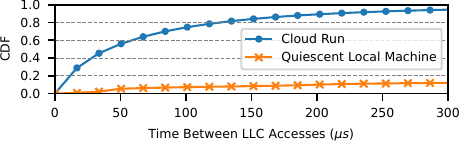}
    \caption{CDF of the time between accesses by background activity to a randomly chosen LLC set.}
    \Description{CDF of the time between accesses by background activity to a randomly chosen LLC set.}
    \label{fig:llc_interval}
    \vspace{2mm}
    \centering
    \includegraphics[width=\linewidth]{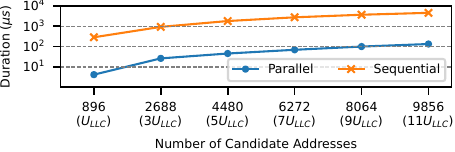}
    \caption{Different $\testEV$'s  execution times on \CR
    under various number of candidate addresses.}
    \Description{Different $\testEV{s}$' execution times on \CR
    for  various numbers of candidate addresses.}
    \label{fig:vwindow}
    \vspace{-2mm}
\end{figure}

\para{Results}
Figure~\ref{fig:llc_interval} shows the cumulative distribution function (CDF)
of the time between LLC accesses by background activity to a randomly chosen LLC set in both environments.
On \CR, the average LLC access rate is \EVResGCPLLCArrival accesses per millisecond per set.
In the local environment with minimal noise,
the average access rate is merely \EVResLocalLLCArrival accesses per millisecond per set.
Figure~\ref{fig:vwindow} shows the execution time of the parallel and sequential $\testEV$,
for varying numbers of candidate addresses on \CR.
As the execution times of $\testEV$ in the local environment follow similar trends,
we omit them in the plot.
It can shown that, on average, the execution times of the sequential and parallel $\testEV$
are \EVResLocSequentialAvgReduction and \EVResLocParallelAvgReduction lower
in the local environment compared to \CR, respectively.

These results explain why existing address pruning algorithms
show unsatisfactory effectiveness on \CR.
For \PrimeScope, when using sequential $\testEV$
to identify the first congruent address,
it is expected to test $11U_{LLC}$ candidate addresses.
This takes approximately \EVResSequentialExpDuration on average.
However, during this time,
the target LLC set is expected to experience \EVResExpDurationNumAcc
background LLC accesses.
Consequently, \PrimeScope's $\testEV$
very likely reports erroneous results under this level of noise.

As for group testing,
its parallel $\testEV$ executes an order of magnitude faster than the sequential $\testEV$.
For example, it takes only \EVResParallelExpDuration to test $11U_{LLC}$ candidates.
Given the background LLC access rate,
the probability of the set \emph{not} being accessed
during the $\testEV$ execution is about \EVResParallelProbNoAccess.
This permits the parallel $\testEV$ a reasonable chance to complete
without experiencing interference from background accesses.
In combination with the backtracking mechanism~\cite{EVSetTheoryPractice},
group testing has a substantially higher probability of success compared to \PrimeScope on \CR.
Still, both \GT and \GTOpt experience a large number of backtracks
due to erroneous $\testEV$ results and are drastically slowed down on \CR.
For example, the optimized \GTOpt performs an average number of
\EVResGCPNoFilterVilaOptBacktracks backtracks per eviction set on \CR,
while it only needs \EVResLocalNoFilterVilaOptBacktracks backtracks
on average in the local environment.

\section{Constructing Eviction Sets in the Cloud}
\label{sec:evset}

Based on the insights from Section~\ref{sec:noise},
we propose two techniques that enable {\em fast} (and therefore also noise-resilient),
eviction set construction in the cloud:
{\em L2-driven candidate address
filtering} (Section~\ref{sec:evset:l2-filtering})
and a {\em Binary Search-based} algorithm for address
pruning (Section~\ref{sec:evset:straw}).

\vspace{-2mm}
\subsection{L2-driven Candidate Address Filtering}
\label{sec:evset:l2-filtering}

To speed-up the eviction set construction,
we propose to reduce the candidate set size with
an algorithm-agnostic optimization that we  call
\emph{candidate address filtering}.
Our insight is that the L2 set index bits are typically a subset of the LLC/SF set index bits.
For example, \SKX uses PA bits 15-6 as the L2 set index and
PA bits 16-6 as the LLC/SF set index (Figure~\ref{fig:skx_pa_mapping}).
Hence, if addresses $A$ and $B$ are not congruent in the L2,
then $A$ and $B$ have different PA bits 15-6 and, therefore,
they must \emph{not} be congruent in the LLC/SF.

Based on this insight,
we introduce a new  {\em candidate filtering  step}
after candidate set construction and before address pruning.
Assume that we want to construct an eviction set for an LLC/SF set to which
an attacker-accessible address $\Taddr$ maps.
To perform the candidate filtering,
we first construct an L2 eviction set for $\Taddr$.
Then, using the L2 eviction set,
we test whether it can evict each
address from the candidate set.
If a candidate address $A$ cannot be evicted by the L2 eviction set,
then it implies that $A$ and $\Taddr$ are
not congruent in either the L2 or the LLC/SF.
Consequently, $A$ is removed from the candidate set.
After candidate filtering,
the candidate set contains only addresses that are congruent with $\Taddr$ in the L2.
These filtered addresses are passed to the address pruning
algorithm to find a minimal LLC/SF eviction set.

As \SKX has an L2 uncertainty of $U_{L2}=\SKXLTwoUncertainty$,
only about $1/\SKXLTwoUncertainty$ of the candidate addresses are congruent with $\Taddr$ in L2.
Therefore, the size of the
filtered candidate set is only about $1/\SKXLTwoUncertainty$ of the original set size.
On a common 28-slice \SKX CPU,
we expect to find one congruent address every $U_{LLC}=\SKXLLCUncertainty$
candidates in the candidate set before filtering.
With candidate filtering,
we now expect to find one congruent address every $896/16=56$ candidates.

Since the candidate set is universally used by different address pruning algorithms,
including both group testing and \PrimeScope,
our candidate filtering is a generic optimization.
Moreover, in modern processors,
the number of L2 sets is typically smaller than the number of LLC sets in one LLC slice.
Hence, the property that the L2 set index bits are a subset of the LLC set index bits
generally holds for other processors as well,
Therefore, the candidate filtering optimization also applies to them.
Lastly, the idea of candidate filtering can be applied to a more restricted environment
where the attacker cannot even control the page offset bits~\cite{EVSetTheoryPractice}.
In such an environment,
the attacker can hierarchically construct L1 and L2 eviction sets
to gradually filter candidates for the next lower cache level.

\vspace{-2mm}
\subsection{Using Binary Search for Address Pruning}
\label{sec:evset:straw}

To further speed-up eviction set construction in the cloud,
we propose a new address pruning algorithm based on binary search.
Our algorithm uses  \emph{parallel} $\testEV$.

\para{Algorithm design}
\label{sec:evset:straw-design}
Given a list of candidate addresses,
we test whether the first $n$ addresses can evict a target address $\Taddr$.
For a $W$-way cache,
increasing $n$ from zero will result in a negative test outcome
until the first $n$ addresses include $W$ congruent addresses.
We define the \emph{tipping point}, denoted by $\tau$,
as the smallest $n$ for which the first $n$ addresses evict $\Taddr$.
Therefore, $\tau$ is the index of the $W$-th congruent address in the list,
assuming that the indexation begins from $1$.
For a given $n$, if the first $n$ addresses   evict $\Taddr$,
it means that $n\geq\tau$; otherwise, $n<\tau$.
Our main idea is to use binary search to efficiently determine $\tau$
and thus identify one congruent address.
Then, we exclude the congruent address from any future search,    %
and repeat the binary search process until $W$ different
congruent addresses are found.

Figure~\ref{lst:straw-algo} shows the pseudo code of the algorithm.
It takes as inputs a target address $T_a$,
an array of addresses $addrs$ representing the candidate set,
and the array size $N$.
The array $addrs$ should contain at least $W$ congruent addresses,
and thus $N \geq W$.
The algorithm iteratively finds $W$ congruent addresses
by finding the tipping point at each iteration
(Lines~\ref{straw:line:for-loop-start}--\ref{straw:line:for-loop-end} in Figure~\ref{lst:straw-algo}).
Within each iteration,
the algorithm tests in a loop if the first $n=\lfloor(\LB+\UB)/2\rfloor$ addresses
from $addrs$ can evict $T_a$ (Line~\ref{straw:line:testev}).
The variables $\LB$ and $UB$ are then updated in a manner that
$\LB$ always represents the largest $n$ such that the first $n$ addresses
\emph{cannot} evict $T_a$ and
$\UB$ always represents the smallest $n$ such that the first $n$ addresses
\emph{can} evict $T_a$.
Therefore, when $\UB=\LB+1$,
$\UB$ is the tipping point of iteration $i$,
denoted by $\tau_i$.
Consequently, the $\tau_i$-th address of the array
is a congruent address.
The algorithm then swaps
the just-found congruent address with the $i$-th address in $addrs$
and proceeds to the next iteration (Line~\ref{straw:line:swap}).

\begin{figure}[t]
    \begin{mdframed}
        \begin{lstlisting}
// T_a: target address
// addrs: an array of candidate addresses
// N: size of the addrs array (N >= W)
size_t LB, UB = N;
for (size_t i = 1; i <= W; i++) { |\label{straw:line:for-loop-start}|
    LB = i - 1; |\label{straw:line:LB-reset}|
    while (UB - LB != 1) {
        n = (LB + UB) / 2;
        if (TestEviction(T_a, addrs, n)) |\label{straw:line:testev}|
            UB = n; // T_a can be evicted
        else
            LB = n; // T_a cannot be evicted
    }
    size_t tau_i = UB;
    swap(addrs[i], addrs[tau_i]); |\label{straw:line:swap}|
} // addrs[1]~addrs[W] form an eviction set |\label{straw:line:for-loop-end}|
        \end{lstlisting}
    \end{mdframed}
    \caption[]{Pseudo code of our proposed algorithm. %
     All array indexes start from $1$.
    Parallel $TestEviction(T_a,\, addrs,\, n)$ returns a boolean value that indicates
    whether the first $n$ candidate addresses from array $addrs$
    can evict the target $\Taddr$.}
    \Description{Pseudo code of our proposed algorithm}
    \label{lst:straw-algo}
    \vspace{-5mm}
\end{figure}

Before the binary search in the next iteration starts,
$\LB$ is reset to $i-1$ (Line~\ref{straw:line:LB-reset}),
as the first $i-1$ addresses are the congruent addresses found in previous iterations
and are thus excluded from the search.
In contrast, $\UB$ needs \emph{not} to be reset to $N$,
as the first $\UB$ addresses always contain $W$ congruent addresses due to the swapping.
Finally, after $W$ iterations,
the first $W$ addresses in $addrs$ form a minimal eviction set for $\Taddr$
(Line~\ref{straw:line:for-loop-end}).

\begin{figure}[t]
    \centering
    \includegraphics[width=.86\linewidth]{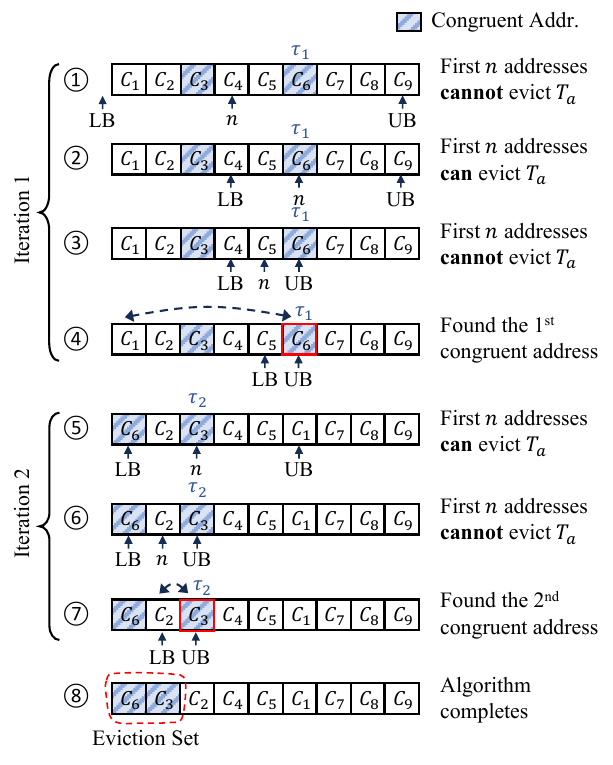}
    \caption{Illustration of our proposed binary search-based algorithm (assuming $W=2$).
    Blocks with shaded pattern represent congruent candidate addresses.
    }
    \Description{Illustration of our proposed algorithm}
    \label{fig:straw-example}
    \vspace{-5mm}
\end{figure}

\para{Example}
Figure~\ref{fig:straw-example} demonstrates the algorithm with  a nine-address
 candidate set ($\Ci{1}, \Ci{2}, \dots, \Ci{9}$)
and a target cache with associativity $W=2$.
Initially, we set $i=1$, $\LB=0$, $\UB=N=9$,
and $n=\lfloor(\UB+\LB)/2\rfloor=4$ (Step \circled{1}).
Because the first $n=4$ addresses cannot evict $\Taddr$,
we set $\LB=n=4$ and update $n$ to $\lfloor(\UB+\LB)/2\rfloor=6$   (Step \circled{2}).
With the updated $n$,
the first $n=6$ addresses now can evict $\Taddr$,
so we set $\UB=n=6$ and update $n=5$ (Step \circled{3}).
This process is repeated until $\UB=\LB+1=6$ (Step \circled{4}).
At this point, $\Ci{6}$ is found to be a congruent address,
which is saved to the front of the list by swapping it with $\Ci{1}$.
Then, we increment $i$ to $2$, set $\LB=i-1=1$ without changing $\UB$ (Step~\circled{5}),
and repeat the binary search (Steps~\circled{5}--\circled{7}).
The algorithm finishes once $W$ congruent addresses are found (Step \circled{8}),
which form a minimal eviction set for $\Taddr$.

\para{Backtracking mechanism}
When the $\testEV$ returns a false-positive result due to environmental noise,
our algorithm can incorrectly set $\UB$ to a value smaller than $\tau$.
As a result, the binary search may incorrectly identify a non-congruent address
as a congruent one.
This erroneous state is detected if the first $\UB$ addresses
cannot evict $\Taddr$ after the binary search for the iteration finishes.
To recover from this state,
we gradually increase $\UB$
with a large stride
until the first $\UB$ addresses can evict $\Taddr$
and restart the binary search.

\para{Comparison to existing algorithms}
Unlike \PrimeScope,
our algorithm uses   parallel $\testEV$.
As discussed in Section~\ref{sec:noise:understand},
parallel $\testEV$ is at least an order of magnitude faster than
sequential $\testEV$. Therefore,
our algorithm is faster %
than \PrimeScope.

Compared to group testing,
both our algorithm and group testing can use parallel $\testEV$.
Assume that we use the number of memory accesses as a proxy for execution time.
Using our algorithm,
it takes $O(\log{N})$ parallel $\testEV$ executions
to find a tipping point, where $N$ is the candidate set size.
Since each parallel $\testEV$ needs to make $O(N)$ memory accesses,
it takes $O(N\log{N})$ accesses to find one congruent address.
As we need to find $W$ congruent addresses,
the end-to-end execution requires $O(WN\log{N})$ accesses.
In contrast,  group testing
requires $O(W^2N)$ accesses.
Therefore, whether group testing or our algorithm makes fewer accesses,
and consequently executes faster,
depends on the specific values of $W$ and $\log{N}$.

As an intuitive comparison,
the ratio of the number of accesses made by
group testing over our algorithm is estimated by $O(W/\log{N})$.
Since we use $N=3UW$ (Section~\ref{sec:noise:ev-fails}),
we rewrite the ratio as $O(W/\log{(UW)})$.
This suggests that in caches with high associativity (i.e., a large $W$),
group testing \emph{tends} to make more accesses than our algorithm.
This is supported by our experiments
in Section~\ref{sec:evset:eval}.

\vspace{-3mm}
\subsection{Evaluating Our Optimizations}
\label{sec:evset:eval}

We evaluate group testing, \PrimeScope, and our binary search-based algorithm
 with candidate filtering
 in both the \CR and local environments.
We use the same methodology as the experiment in Section~\ref{sec:noise:ev-fails},
except for reducing the time limit of constructing one eviction set to \EVConfigSingleEVTimeout (because of candidate
filtering).
Each algorithm is evaluated in three scenarios:
(1) \SingleSet, where we construct a single eviction set for a randomly chosen SF set;
(2) \PageOffset, where we construct eviction sets for all SF sets at a randomly chosen page offset;
and (3) \WholeSF, where we construct eviction sets for all SF sets in the system.
Our experiments include \EVConfigGCPTotalNumSingleSetMeasures and
\EVConfigLocalSingleEVMeasures measurements per algorithm
in the cloud and local environments, respectively, in \SingleSet;
\EVConfigGCPTotalNumPageOffsetMeasures and
\EVConfigLocalPageOffsetMeasures in \PageOffset;
and \EVConfigGCPTotalNumWholeSetMeasures and
\EVConfigLocalWholeSetMeasures in \WholeSF.

\begin{table*}[ht]
    \caption{Eviction set construction effectiveness of various algorithms under different configurations. The number of eviction sets may vary between local and cloud because the experiments use machines with different number of slices.
    }
    \label{tab:evres}
    \centering
    \small
    \setlength{\tabcolsep}{4pt}
    \begin{tabular}{|c|c|cccc|cccc|cccc|}
        \hline
        \multirow{2}{*}{\textbf{Env.}} &
          \multirow{2}{*}{\textbf{Metrics}} &
          \multicolumn{4}{c|}{\begin{tabular}[c]{@{}c@{}}\textbf{\SingleSet}\\
                                {\footnotesize \# Ev sets: Local=1, Cloud=1}\end{tabular}} &
          \multicolumn{4}{c|}{\begin{tabular}[c]{@{}c@{}}\textbf{\PageOffset}\\
                                {\footnotesize \# Ev sets: Local=$704$, Cloud=$896$}\end{tabular}} &
          \multicolumn{4}{c|}{\begin{tabular}[c]{@{}c@{}}\textbf{\WholeSF}\\
                                {\footnotesize \# Ev sets: Local=\SKXLocalLLCSetsTotal, Cloud=\SKXLLCSetsTotal}\end{tabular}} \\ \cline{3-14}
         & \rule{0pt}{2.2ex}
           &
          \multicolumn{1}{c|}{\textbf{\GT}} &
          \multicolumn{1}{c|}{\textbf{\GTOpt}} &
          \multicolumn{1}{c|}{\textbf{\PSBest}} &
          \textbf{\Ours} &
          \multicolumn{1}{c|}{\textbf{\GT}} &
          \multicolumn{1}{c|}{\textbf{\GTOpt}} &
          \multicolumn{1}{c|}{\textbf{\PSBest}} &
          \textbf{\Ours} &
          \multicolumn{1}{c|}{\textbf{\GT}} &
          \multicolumn{1}{c|}{\textbf{\GTOpt}} &
          \multicolumn{1}{c|}{\textbf{\PSBest}} &
          \textbf{\Ours} \\ \hline
        \multirow{4}{*}{\begin{tabular}[c]{@{}c@{}}Quiescent\\Local\end{tabular}} &
          Succ. Rate \rule{0pt}{2.2ex} &
          \multicolumn{1}{c|}{\EVResLocalFilterVilaSuccRate} &
          \multicolumn{1}{c|}{\EVResLocalFilterVilaOptSuccRate} &
          \multicolumn{1}{c|}{\EVResLocalFilterPSOptSuccRate} &
                              \EVResLocalFilterStrawSuccRate &
          \multicolumn{1}{c|}{\EVResLocalPageOffsetVilaAvgSFSuccRate} &
          \multicolumn{1}{c|}{\EVResLocalPageOffsetVilaOptAvgSFSuccRate} &
          \multicolumn{1}{c|}{\EVResLocalPageOffsetPSOptAvgSFSuccRate} &
                              \EVResLocalPageOffsetStrawAvgSFSuccRate &
          \multicolumn{1}{c|}{\EVResLocalWholeSetVilaAvgSFSuccRate} &
          \multicolumn{1}{c|}{\EVResLocalWholeSetVilaOptAvgSFSuccRate} &
          \multicolumn{1}{c|}{\EVResLocalWholeSetPSAvgSFSuccRate} &
                              \EVResLocalWholeSetStrawAvgSFSuccRate \\ \cline{2-14}
         &
          Avg. Time \rule{0pt}{2.2ex} &
          \multicolumn{1}{c|}{\EVResLocalFilterVilaAvgDuration} &
          \multicolumn{1}{c|}{\EVResLocalFilterVilaOptAvgDuration} &
          \multicolumn{1}{c|}{\EVResLocalFilterPSOptAvgDuration} &
                              \EVResLocalFilterStrawAvgDuration &
          \multicolumn{1}{c|}{\EVResLocalPageOffsetVilaAvgDuration} &
          \multicolumn{1}{c|}{\EVResLocalPageOffsetVilaOptAvgDuration} &
          \multicolumn{1}{c|}{\EVResLocalPageOffsetPSOptAvgDuration} &
                              \EVResLocalPageOffsetStrawAvgDuration &
          \multicolumn{1}{c|}{\EVResLocalWholeSetVilaAvgDuration} &
          \multicolumn{1}{c|}{\EVResLocalWholeSetVilaOptAvgDuration} &
          \multicolumn{1}{c|}{\EVResLocalWholeSetPSAvgDuration} &
                              \EVResLocalWholeSetStrawAvgDuration \\ \cline{2-14}
         &
          Stddev Time \rule{0pt}{2.2ex} &
          \multicolumn{1}{c|}{\EVResLocalFilterVilaStdDuration} &
          \multicolumn{1}{c|}{\EVResLocalFilterVilaOptStdDuration} &
          \multicolumn{1}{c|}{\EVResLocalFilterPSOptStdDuration} &
                              \EVResLocalFilterStrawStdDuration &
          \multicolumn{1}{c|}{\EVResLocalPageOffsetVilaStdDuration} &
          \multicolumn{1}{c|}{\EVResLocalPageOffsetVilaOptStdDuration} &
          \multicolumn{1}{c|}{\EVResLocalPageOffsetPSOptStdDuration} &
                              \EVResLocalPageOffsetStrawStdDuration &
          \multicolumn{1}{c|}{\EVResLocalWholeSetVilaStdDuration} &
          \multicolumn{1}{c|}{\EVResLocalWholeSetVilaOptStdDuration} &
          \multicolumn{1}{c|}{\EVResLocalWholeSetPSStdDuration} &
                              \EVResLocalWholeSetStrawStdDuration \\ \cline{2-14}
         &
          Med. Time \rule{0pt}{2.2ex} &
          \multicolumn{1}{c|}{\EVResLocalFilterVilaMedianDuration} &
          \multicolumn{1}{c|}{\EVResLocalFilterVilaOptMedianDuration} &
          \multicolumn{1}{c|}{\EVResLocalFilterPSOptMedianDuration} &
                              \EVResLocalFilterStrawMedianDuration &
          \multicolumn{1}{c|}{\EVResLocalPageOffsetVilaMedianDuration} &
          \multicolumn{1}{c|}{\EVResLocalPageOffsetVilaOptMedianDuration} &
          \multicolumn{1}{c|}{\EVResLocalPageOffsetPSOptMedianDuration} &
                              \EVResLocalPageOffsetStrawMedianDuration &
          \multicolumn{1}{c|}{\EVResLocalWholeSetVilaMedianDuration} &
          \multicolumn{1}{c|}{\EVResLocalWholeSetVilaOptMedianDuration} &
          \multicolumn{1}{c|}{\EVResLocalWholeSetPSMedianDuration} &
                              \EVResLocalWholeSetStrawMedianDuration \\ \hline\hline
        \multirow{4}{*}{\begin{tabular}[c]{@{}c@{}}Cloud\\ Run\end{tabular}} &
        Succ. Rate \rule{0pt}{2.2ex} &
        \multicolumn{1}{c|}{\EVResGCPFilterVilaSuccRate} &
        \multicolumn{1}{c|}{\EVResGCPFilterVilaOptSuccRate} &
        \multicolumn{1}{c|}{\EVResGCPFilterPSOptSuccRate} &
                            \EVResGCPFilterStrawSuccRate &
        \multicolumn{1}{c|}{\EVResGCPPageOffsetVilaAvgSFSuccRate} &
        \multicolumn{1}{c|}{\EVResGCPPageOffsetVilaOptAvgSFSuccRate} &
        \multicolumn{1}{c|}{\EVResGCPPageOffsetPSAvgSFSuccRate} &
                            \EVResGCPPageOffsetStrawAvgSFSuccRate &
        \multicolumn{1}{c|}{\EVResGCPWholeSetVilaAvgSFSuccRate} &
        \multicolumn{1}{c|}{\EVResGCPWholeSetVilaOptAvgSFSuccRate} &
        \multicolumn{1}{c|}{\EVResGCPWholeSetPSAvgSFSuccRate} &
                            \EVResGCPWholeSetStrawAvgSFSuccRate \\ \cline{2-14}
       &
        Avg. Time \rule{0pt}{2.2ex} &
        \multicolumn{1}{c|}{\EVResGCPFilterVilaAvgDuration} &
        \multicolumn{1}{c|}{\EVResGCPFilterVilaOptAvgDuration} &
        \multicolumn{1}{c|}{\EVResGCPFilterPSOptAvgDuration} &
                            \EVResGCPFilterStrawAvgDuration &
        \multicolumn{1}{c|}{\EVResGCPPageOffsetVilaAvgDuration} &
        \multicolumn{1}{c|}{\EVResGCPPageOffsetVilaOptAvgDuration} &
        \multicolumn{1}{c|}{\EVResGCPPageOffsetPSAvgDuration} &
                            \EVResGCPPageOffsetStrawAvgDuration &
        \multicolumn{1}{c|}{\EVResGCPWholeSetVilaAvgDuration} &
        \multicolumn{1}{c|}{\EVResGCPWholeSetVilaOptAvgDuration} &
        \multicolumn{1}{c|}{\EVResGCPWholeSetPSAvgDuration} &
                            \EVResGCPWholeSetStrawAvgDuration \\ \cline{2-14}
       &
        Stddev Time \rule{0pt}{2.2ex} &
        \multicolumn{1}{c|}{\EVResGCPFilterVilaStdDuration} &
        \multicolumn{1}{c|}{\EVResGCPFilterVilaOptStdDuration} &
        \multicolumn{1}{c|}{\EVResGCPFilterPSOptStdDuration} &
                            \EVResGCPFilterStrawStdDuration &
        \multicolumn{1}{c|}{\EVResGCPPageOffsetVilaStdDuration} &
        \multicolumn{1}{c|}{\EVResGCPPageOffsetVilaOptStdDuration} &
        \multicolumn{1}{c|}{\EVResGCPPageOffsetPSStdDuration} &
                            \EVResGCPPageOffsetStrawStdDuration &
        \multicolumn{1}{c|}{\EVResGCPWholeSetVilaStdDuration} &
        \multicolumn{1}{c|}{\EVResGCPWholeSetVilaOptStdDuration} &
        \multicolumn{1}{c|}{\EVResGCPWholeSetPSStdDuration} &
                            \EVResGCPWholeSetStrawStdDuration \\ \cline{2-14}
       &
        Med. Time \rule{0pt}{2.2ex} &
        \multicolumn{1}{c|}{\EVResGCPFilterVilaMedianDuration} &
        \multicolumn{1}{c|}{\EVResGCPFilterVilaOptMedianDuration} &
        \multicolumn{1}{c|}{\EVResGCPFilterPSOptMedianDuration} &
                            \EVResGCPFilterStrawMedianDuration &
        \multicolumn{1}{c|}{\EVResGCPPageOffsetVilaMedianDuration} &
        \multicolumn{1}{c|}{\EVResGCPPageOffsetVilaOptMedianDuration} &
        \multicolumn{1}{c|}{\EVResGCPPageOffsetPSMedianDuration} &
                            \EVResGCPPageOffsetStrawMedianDuration &
        \multicolumn{1}{c|}{\EVResGCPWholeSetVilaMedianDuration} &
        \multicolumn{1}{c|}{\EVResGCPWholeSetVilaOptMedianDuration} &
        \multicolumn{1}{c|}{\EVResGCPWholeSetPSMedianDuration} &
                            \EVResGCPWholeSetStrawMedianDuration \\ \hline
    \end{tabular}
    \vspace{-3mm}
\end{table*}

\para{Results}
Table~\ref{tab:evres} lists the success rate and execution time of each algorithm
under different scenarios in both the \CR and local environments.
The execution time measures \emph{both candidate filtering and address pruning}.
As we find that \PS and \PSOpt have similar success rates and execution times
after applying candidate filtering, Table~\ref{tab:evres}
 only shows the one with the shortest average execution time, and calls it \PSBest. We call
our binary search-based algorithm \Ours.

The \SingleSet scenario in Table~\ref{tab:evres} is directly comparable to the scenario
in Table~\ref{tab:eff_nofilter}. Table~\ref{tab:evres} shows the effectiveness of candidate filtering on \CR,
as it leads to significantly shortened execution times.
For example, the average execution time of \GTOpt is reduced from
\EVResGCPNoFilterVilaOptAvgDuration to \EVResGCPFilterVilaOptAvgDuration.
The resulting success rate also increases substantially.
Indeed, for \GTOpt, it goes from
\EVResGCPNoFilterVilaOptSuccRate to \EVResGCPFilterVilaOptSuccRate.

Recall that the average execution time comprises both candidate filtering
and addresses pruning. In the \SingleSet scenario, it can be shown that
candidate filtering on \CR takes on average \GCPLTwoFilterSingleSetAvgDuration,
which dominates the execution time.
As a result, the average execution times are similar across all algorithms.
As will be shown in Section~\ref{sec:cand-filter-overhead},
the portion of the execution time spent on candidate filtering
drastically decreases when building numerous eviction sets
in the \PageOffset and \WholeSF scenarios.

Next, consider \PageOffset.
All the algorithms experience increases in average execution times as they go from the
local to the \CR environments.
Comparing group testing to our algorithm on \CR, we see that
\GT and \GTOpt take \EVResGCPPageOffsetVilaDurationOverhead and
\EVResGCPPageOffsetVilaOptDurationOverhead more time to build eviction sets on average,
as we find that \GT and \GTOpt make
\EVResGCPPageOffsetVilaAccOverhead and \EVResGCPPageOffsetVilaOptAccOverhead
more memory accesses than \Ours.
As for \PSBest,
it takes on average \EVResGCPPageOffsetPSDurationOverhead more time than \Ours,
due to its use of the sequential $\testEV$.

The results for \WholeSF are qualitatively similar to \PageOffset,
except for larger drops in success rates as we go from the
local to the \CR environments.
Still, while the average success rates of \GT, \GTOpt, \PSBest, and \Ours on \CR
are \EVResGCPWholeSetVilaAvgSFSuccRate, \EVResGCPWholeSetVilaOptAvgSFSuccRate,
\EVResGCPWholeSetPSAvgSFSuccRate, and \EVResGCPWholeSetStrawAvgSFSuccRate, respectively,
the medians are \EVResGCPWholeSetVilaMedianSFSuccRate,
\EVResGCPWholeSetVilaOptMedianSFSuccRate, \EVResGCPWholeSetPSMedianSFSuccRate,
and \EVResGCPWholeSetStrawMedianSFSuccRate, respectively.

To summarize,
the combination of candidate filtering and our binary search-based algorithm
offers significant performance improvements
over the \emph{well-optimized} state-of-the-art algorithms.
On \CR, they reduce the time to construct eviction sets for all SF sets in the system
from an expected duration of \EVResGCPNoFilterVilaOptDurationTotal (Section~\ref{sec:noise:ev-fails})
to a mere \EVResGCPWholeSetStrawAvgDurationMin (last column of Table~\ref{tab:evres}),
with a median success rate of \EVResGCPWholeSetStrawMedianSFSuccRate.
These improvements
make the LLC \PrimeProbe attack in the cloud feasible.

\vspace{-5mm}
\subsubsection{Overhead of Candidate Filtering}
\label{sec:cand-filter-overhead}

As indicated before, it takes \GCPLTwoFilterSingleSetAvgDuration
to complete one candidate filtering on \CR.
This time includes constructing one L2 eviction set
and using it to filter candidates.
While this time dominates the execution time when constructing
a \emph{single} eviction set (Section~\ref{sec:evset:eval}),
the same filtered candidates can be reused to construct many eviction sets
for LLC/SF sets that are mapped to the same L2 set.
For example, in the 28-slice \SKX processor used in our \CR evaluation,
constructing the \SKXLLCUncertainty LLC/SF sets in the \PageOffset scenario
requires only $16$ candidate filtering executions,
which takes \EVResGCPCandsFilter on average.
This execution time makes up a small portion of
the total execution time in \PageOffset
(\EVResGCPPageOffsetStrawAvgDuration in Table~\ref{tab:evres}).

In the \WholeSF scenario,
a naive process would build eviction sets for all \SKXLTwoSets L2 sets
and execute candidate filtering \SKXLTwoSets times.
We optimize the process by exploiting the following property of the L2:
if addresses $A$ and $B$ are congruent in L2,
then $A'=A+\delta$ and $B'=B+\delta$ are also congruent in L2---as long as
the $\delta$ is small enough such that $A$ and $A'$ belong to the same page,
and $B$ and $B'$ belong to the same page~\cite{oren2015spy,LLCPractical,ssa}.

Exploiting this property,
we first construct $16$ eviction sets for all L2 sets at page offset \texttt{0x0}.
Then, we use each eviction set to generate a filtered candidate set
at page offset \texttt{0x0}.
Finally, we can derive a new filtered candidate set at page offset $\delta$
by adding $\delta$ to each candidate address
of the filtered candidate set at page offset \texttt{0x0}.
As a result, the \WholeSF scenario requires only
$16$ L2 eviction set constructions and candidate filtering executions.
The time of completing candidate filtering (\EVResGCPCandsFilter)
is negligible compared to the total execution time in \WholeSF
(\EVResGCPWholeSetStrawAvgDuration in Table~\ref{tab:evres}).

\vspace{-4mm}
\subsubsection{Other Intel Server Platforms and Target Caches}
\label{sec:other-intel-cpus}

As discussed in Section~\ref{sec:evset:straw},
group testing tends to incur a higher execution overhead over our binary search-based algorithm
when the cache associativity increases.
To illustrate this trend,
we measure the performance of eviction set construction on \ICX, which
features caches with higher associativity than in \SKX.
Specifically, \ICX has a 16-way SF and a 20-way L2 cache,
whereas \SKX has a 12-way SF and a 16-way L2 cache.

Because we do not see \ICX being used on \CR,
we measure the performance on local quiescent \SKX and \ICX machines. %
The \SKX machine utilized is the same as in prior experiments.
The \ICX machine uses an Intel Xeon Gold 5320,
which has 26 LLC/SF slices.
For each machine and algorithm,
we measure the  time to construct  a single SF or L2 eviction set
$\OCConfigRepeats$ times.
Candidate filtering is enabled for SF eviction set construction,
but its time is \emph{not} included in our measurements.

First, we consider constructing eviction sets for the SF.
\GT, \GTOpt, and \Ours take, on average,
\EVResLocalFilterVilaAvgDurationNoCandsFilter,
\EVResLocalFilterVilaOptAvgDurationNoCandsFilter,
and \EVResLocalFilterStrawAvgDurationNoCandsFilter, respectively,
to construct a single eviction set for the 12-way SF of \SKX.
The same process takes \GT, \GTOpt, and \Ours on average
\OCResICXSFVilaAvgDura, \OCResICXSFVilaOptAvgDura,
and \OCResICXSFStrawAvgDura, respectively, for the 16-way SF of \ICX.
As we go from \SKX to \ICX, the ratio \GT/\Ours and \GTOpt/\Ours changes from
1.91 and 1.51 to 2.27 and 1.83, respectively.

Similarly,
\GT, \GTOpt, and \Ours take, on average,
\OCResSKXLTwoVilaAvgDura, \OCResSKXLTwoVilaOptAvgDura,
and \OCResSKXLTwoStrawAvgDura, respectively,
to construct a single eviction set for the 16-way L2 of \SKX.
The same process takes \GT, \GTOpt, and \Ours on average
\OCResICXLTwoVilaAvgDura, \OCResICXLTwoVilaOptAvgDura,
and \OCResICXLTwoStrawAvgDura, respectively, for the 20-way L2 of \ICX.
As we go from \SKX to \ICX, the ratio \GT/\Ours and \GTOpt/\Ours changes from
1.87 and 1.43 to 6.35 and 3.58, respectively.

\vspace{-2mm}

\section{Monitoring Memory Accesses \& Identifying Target Cache Sets}
\label{sec:psd}

Eviction set construction is the first step of
an end-to-end LLC attack (\AttackStep{1} in Table~\ref{tab:attack-steps}).
In this section, we improve the remaining steps with two new techniques.
First, Section~\ref{sec:psd:monitor} introduces {\em Parallel Probing}, which enables the
monitoring of victim memory accesses with
high time resolution. This technique optimizes \AttackStep{2}
(identify target sets) and \AttackStep{3} (exfiltrate
information) in Table~\ref{tab:attack-steps}
for the noisy cloud environment.
Second, Section~\ref{sec:psd:psd} leverages {\em Power Spectral Density}~\cite{stoica2005spectral} from signal processing to
easily identify the victim's target cache set. This technique optimizes
\AttackStep{2} in Table~\ref{tab:attack-steps}
for the noisy cloud environment.

\vspace{-2mm}
\subsection{Parallel Probing for Memory Access Monitoring}
\label{sec:extraction:monitor}
\label{sec:psd:monitor}

Given a cache set to monitor, the attacker can detect memory accesses to that set with \PrimeProbe~(Section~\ref{sec:back:cache-side-channel}).
It is vital that \emph{both} prime and probe latencies are short.
A short probe latency enables the attacker to monitor
when accesses occurs at a high time resolution~\cite{primescope}.
A short prime latency allows the attacker to quickly prepare
the monitored cache set for detecting the next access.
In a noisy cloud environment,
where a cache set may be frequently accessed by processes of other tenants,
failure to prime the set in a timely manner can
increase the chance of missing the victim's accesses.

To minimize the probe latency,
\PrimeScope~\cite{primescope} primes a specific line from the eviction set
to become the \emph{eviction candidate (EVC)},
which is the  line to be evicted when a new line needs to be inserted into the set.
This method enables the attacker to check only if the EVC remains cached.
Further, since the EVC can be cached in L1, the probe latency becomes minimal,
leading to a high time resolution.
However, this comes at the cost of using a slower
and more complex priming pattern to prepare the replacement state~\cite{primescope},
which can reduce monitoring effectiveness in a noisy environment.

\parans{Our solution}
We discover that,
due to the high memory-level parallelism supported by modern processors,
simply probing {\em with overlapped accesses}
all the $W$ lines of a minimal eviction set (Section~\ref{sec:back:evset})
results in a probe latency only slightly higher than that of \PrimeScope.
The advantage of this \emph{parallel probing} method is that
it allows us to
prime the cache set without
preparing any replacement state.
Therefore,
parallel probing works irrespective of the replacement policy used by the target cache,
which can be unknown or quite complex~\cite{EVSetTheoryPractice,
qureshi2007adaptive,RRIP,wong2013IvyBridge}.

\para{Evaluating Parallel Probing}
We conduct a covert-channel experiment similar to the one done by Purnal~et~al.~\cite{primescope}
to evaluate two different \PrimeScope strategies and our parallel probing.
In the experiment,
we create a sender and a receiver thread that agree on a target SF set.
The sender thread accesses the target set at a fixed time interval,
while the receiver thread uses \PrimeScope or parallel probing to detect accesses to the target set.
For a sender's access issued at time $t$,
if the receiver detects an access at time $t'\in(t, t+\epsilon)$,
where $\epsilon$ is an error bound,
we say that the sender's access is detected by the receiver.
We use $\epsilon=\CovertConfigErrBnd$ (or \CovertConfigErrBndns).

We conduct this experiment on \CR with varying access intervals.  %
In each experiment,
the sender thread accesses the target SF set \CovertConfigNumEmits times.
We measure the percentage of the sender's accesses
that are detected by the receiver---i.e., the \emph{detection rate}.
We also collect the probe and prime latencies and exclude
outliers that are above \CovertConfigCtxSwitchThresh cycles,
as an interrupt or context switch likely occurred during the operation.
The experiment is done on different hosts on different days
and at different times of day.
We repeat the experiment \CovertConfigNumRepeat times on each host,
totaling \CovertConfigGCPNumMeasure measurements.

For \PrimeScope, we evaluate two prime strategies
discussed by Purnal~et~al.~\cite{primescope}.
The first strategy (\PSFlush)  is to load, flush, and sequentially reload the eviction set.
The second strategy (\PSAlt) is to perform an alternating pointer-chase using \emph{two} eviction sets.
More details of these strategies is found in~\cite{primescope}.
For our parallel probing technique (\ParaProbe),
we use a prime strategy that simply traverses the eviction set $12$ times
with overlapped accesses.

Table~\ref{tab:covert-comp} lists the prime and probe latencies of each  strategy.
The table reveals that the average probe latency of \ParaProbe
is only \CovertResParaResolutionDiff cycles higher than that of \PrimeScope,
yet \ParaProbe exhibits a substantially lower prime latency.

\begin{table}[t]
    \centering
    \caption{Prime and probe latencies of two \PrimeScope strategies and parallel probing on \CR.
             The host processors' frequency is \GCPSKXClockSpeed.
             }
    \label{tab:covert-comp}
    \small
    \setlength{\tabcolsep}{5pt}
    \begin{tabular}{|c|c|c|}
    \hline
    \textbf{Strategy} &
    \begin{tabular}[c]{@{}c@{}}\textbf{Prime Latency}\\(mean $\pm$ std. deviation)\end{tabular} &
    \begin{tabular}[c]{@{}c@{}}\textbf{Probe Latency}\\(mean $\pm$ std. deviation)\end{tabular} \\ \hline
    \PSFlush   & \CovertResPSFlushPrimeLat $\pm$ \CovertResPSFlushPrimeLatStd cycles
               & \multirow{2}{*}{\CovertResPsResolution $\pm$ \CovertResPsResolutionStd cycles} \\ \cline{1-2}
    \PSAlt     & \CovertResPSAltPrimeLat $\pm$ \CovertResPSAltPrimeLatStd cycles
               &  \\ \hline
    \ParaProbe & \CovertResParallelPrimeLat $\pm$ \CovertResParallelPrimeLatStd cycles
               & \CovertResParaResolution $\pm$ \CovertResParaResolutionStd cycles \\ \hline
    \end{tabular}
    \vspace{-4mm}
\end{table}

\begin{figure}[t]
    \centering
    \includegraphics[width=.9\linewidth]{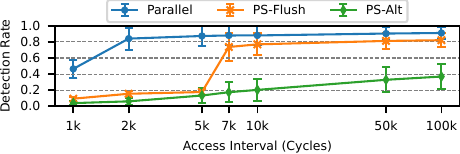}
    \caption{Detection rate of each monitoring strategy
             with various access interval.
             The x-axis employs a logarithmic scale.
             The error bars represent the standard deviations.}
    \Description{Detection rate of each monitoring strategy
                 with various access interval.}
    \label{fig:covert}
    \vspace{-5mm}
\end{figure}

The benefit of this reduced prime latency is depicted in Figure~\ref{fig:covert},
which shows the average detection rate for different access intervals.
With a 2k-cycle access interval,
\ParaProbe achieves an average detection rate of \CovertResParallelTwoKAvgDetectionRate,
while \PSFlush and \PSAlt reach average detection rates of
\CovertResPSFlushTwoKAvgDetectionRate and \CovertResPSAltTwoKAvgDetectionRate, respectively.
The low detection rates of   \PSFlush and \PSAlt are
primarily due to their long prime latencies.

Even when the access interval is sufficiently long
for all strategies to complete priming,
\ParaProbe still maintains the highest detection rate.
With a 100k-cycle access interval, \ParaProbe, \PSFlush, and \PSAlt
attain average detection rates of \CovertResParallelHundredKAvgDetectionRate,
\CovertResPSFlushHundredKAvgDetectionRate,
and \CovertResPSAltHundredKAvgDetectionRate, respectively.
To understand why,
we  inspected a random subset of the detected memory access traces.
In  \PSFlush,
we observe that missed detections mainly result
from noisy accesses made by other tenants to the monitored cache set,
occurring just before the sender's access.
After the receiver detects the noisy access, it
is unable to finish priming before the sender accesses the set.

In \PSAlt, although the receiver initially detects the sender's accesses,
it often later fails to prime the monitored line as the EVC, %
leading to many missed detections.
We believe this might be due to the SF replacement states being
altered by background accesses,
resulting in  failing to prepare the EVC.

\vspace{-1mm}
\subsection{Power Spectral Density for Set Identification}
\label{sec:psd:psd}

To identify the target cache sets (\AttackStep{2} in Table~\ref{tab:attack-steps}),
the attacker can collect a short memory access trace
from each potential target cache set
\emph{while the victim is executing}.
The attacker then applies signal processing techniques
to determine whether a given memory access trace
has any characteristic
that resembles what is expected from a given target cache set.
Prior work has considered characteristics such as
the number of accesses in the trace or the access pattern~\cite{LLCPractical,ssa}.
These characteristics can be hard to identify in the cloud
due to the high level of environmental noise.

\parans{Our solution}
Our insight is that a victim program's accesses to the target cache set
are often periodic in a way that the attacker expects, while this is not
the case for the background accesses.
Therefore, we propose to process the access traces in the frequency domain,
where it is easier to spot the expected periodic patterns.
Specifically,
we estimate the {\em Power Spectral Density} (PSD)~\cite{stoica2005spectral}
of each memory access trace using Welch's method~\cite{welch1967use}.
PSD measures the ``strength'' of the signal at different frequencies~\cite{stoica2005spectral}.
If the access trace is collected from the target set
where the victim makes periodic accesses,
we will observe peaks in the trace's PSD around the expected victim-access frequencies.
If, instead,  the trace is not collected from the target set, it
will have a PSD without the expected peaks.

\parans{Example}
To demonstrate our proposal,
we collect an access trace from a target SF set of a victim program
and another trace from a non-target SF set, and compare the PSD of both traces.
In this example, the victim executes an ECDSA implementation~\cite{mont-ossl}
that will be described in Section~\ref{sec:extraction:ecdsa}.
In this implementation, the victim  processes each individual secret bit in a loop.
The victim accesses the target SF set when an iteration starts
and, if the secret bit being processed in the iteration is zero,
it also accesses the set in the midpoint of the iteration.
The execution of each iteration takes a mostly fixed time duration
of about \MontAvgIterationDuration on a \GCPSKXClockSpeed \SKX machine on \CR.
Because of the access that may occur in the midpoint of an iteration,
the victim's accesses to the target set have a period of about \MontAvgHalfIterationDuration.
Therefore,
we expect to observe a peak in the PSD at the  frequency of
$f=\GCPSKXClockSpeed/\MontAvgHalfIterationDurationNoUnit\approx \MontAvgPSDBaseFrequency$.

\begin{figure}[t]
    \centering
    \begin{subfigure}{.9\linewidth}
        \centering
        \includegraphics[width=\linewidth]{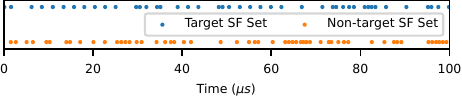}
        \vspace{-3mm}
    \end{subfigure}
    \begin{subfigure}{0.9\linewidth}
        \centering
        \includegraphics[width=\linewidth]{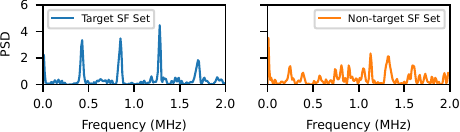}
    \end{subfigure}
    \caption{
    The top plot shows traces of memory access to
    the target SF set (top trace) and the non-target SF set (bottom trace) collected on \CR.
    The two bottom plots show
    the  power spectral density of the two traces.
    }
    \Description{Demonstration of the PSD method with memory access traces
    collected on \CR.}
    \label{fig:mont-psd}
    \vspace{-5mm}
\end{figure}

The top plot of Figure~\ref{fig:mont-psd} shows two \SI{100}{\micro\second}
memory access traces collected on a \GCPSKXClockSpeed \SKX machine on \CR.
The blue dots at the top are the observed accesses to the target SF set;
while the orange dots at the bottom are the observed accesses to the non-target SF set.
For both traces, we see similar numbers of accesses:
$50$ accesses to the target set and $48$ to the non-target set.
It is  difficult to interpret these two patterns.

\begin{figure}
    \centering
    \begin{subfigure}{0.59\linewidth}
        \centering
        \includegraphics[width=.9\linewidth]{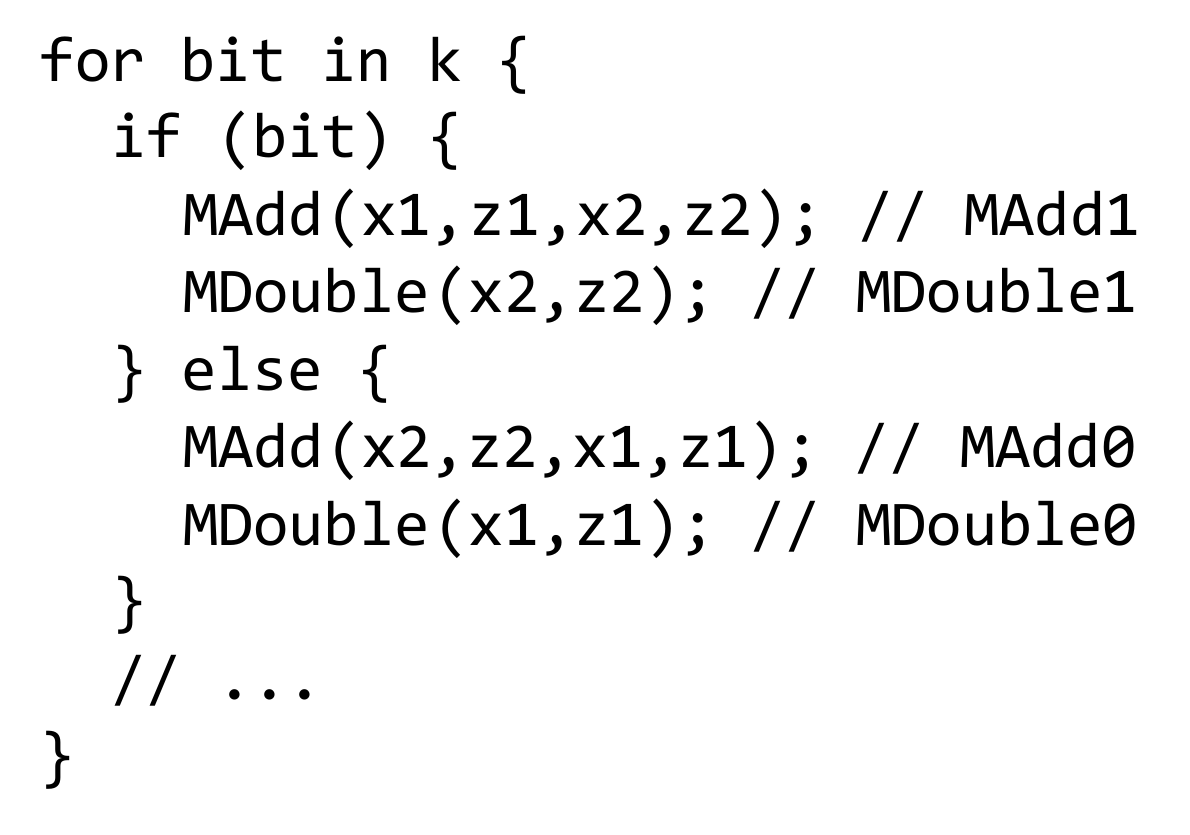}
        \caption{Simplified code snippet.}
        \label{fig:mont-snippet}
    \end{subfigure}
    \begin{subfigure}{0.4\linewidth}
        \centering
        \includegraphics[width=.8\linewidth]{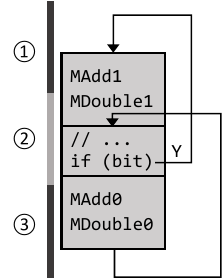}
        \caption{Memory layout.}
        \label{fig:mont-layout}
    \end{subfigure}
    \caption{Simplified vulnerable code snippet (left) and
    its memory layout in VA space (right).
    Each thick vertical line represents a cache line.
    The control-flow edge that exits the loop is omitted
    in the right figure.}
    \Description{Simplified vulnerable code snippet and
                 its memory layout in VA space.}
    \label{fig:mont-code}
    \vspace{-5mm}
\end{figure}

The bottom plots
show the PSD of the access traces collected from the target set (left) and
the non-target set (right).
In the PSD for the target set,
we  clearly see a peak at the base frequency $f=\MontAvgPSDBaseFrequency$ and at multiples of $f$.
In contrast,
in the PSD for the non-target set,
we see no significant peaks at the expected frequency.

\vspace{-1mm}
\section{Demonstrating an End-to-End Attack}
\label{sec:extraction}

In this section,
we demonstrate the combination of our techniques discussed in Sections~\ref{sec:evset}~and~\ref{sec:psd}
by mounting an end-to-end, cross-tenant attack in \CR.
Our demonstration uses
a vulnerable implementation of
Elliptic Curve Digital Signature Algorithm (ECDSA)~\cite{ECDSA}
from OpenSSL 1.0.1e~\cite{mont-ossl}
as an example victim.
While this implementation is deprecated,
we use it solely as a vehicle to illustrate our techniques.

\vspace{-1mm}
\subsection{Attack Outline}
\label{sec:extraction:ecdsa}

The vulnerable ECDSA implementation that we target uses
the Montgomery ladder technique~\cite{joye2002montgomery}
to compute on the nonce $k$,
an ephemeral key that changes with each signing.
The attacker can derive the private key used for signing
by extracting some bits of $k$ across multiple signing operations~\cite{nguyen2003insecurity,mont_flush_reload,
fan2016attacking, de2020tale, aranha2020ladderleak, generalizedECDSAAttack,howgrave2001lattice}.
Thus, the attacker's goal is to learn as many bits of $k$ as possible.
Our demonstration targets curve \texttt{sect571r1},
which uses a 571-bit nonce.

Similar to prior work~\cite{LLCPractical,mont_flush_reload,inci2015seriously},
we assume the attacker knows the memory layout of the library used by the victim.
This assumption generally holds, as victims often install and use
libraries whose binaries are publicly released.
Moreover, as we are targeting a victim web service (Section~\ref{sec:threat}),
we assume the library is loaded once at the victim container startup time and
uses the same VA-PA mapping throughout the container's lifetime.

Figure~\ref{fig:mont-snippet} shows a simplified version of
the   Montgomery ladder implementation~\cite{mont-ossl} that we are targeting.
The code iterates through each bit of the nonce $k$ and
 calls functions \MAdd and \MDouble with different arguments
depending on the value of the bit.
This implementation is resilient to end-to-end timing, as it
executes the same sequence of operations regardless of the bit value.
However, it has secret-dependent control flow.
Since each side of the branch resides on a different cache line,
the program fetches different cache lines based on the value of the nonce bit.
As a result,
the attacker can infer each individual nonce bit by monitoring
code fetch accesses to these cache lines tracked by the SF.

Figure~\ref{fig:mont-layout} shows the memory layout of
the vulnerable code snippet in VA space,
compiled with the default build options and static linkage.
Each thick vertical line represents a different cache line.
Given this layout,
one approach is to monitor accesses to cache line \circled{2}.
Line \circled{2} is used by the \texttt{if} statement,
which is executed at the beginning of an iteration.
As a result, the code fetch accesses made by the \texttt{if} statement
serve as a ``clock'' and mark the iteration boundaries.

Cache line \circled{2} is also utilized by the true direction of the branch.
When the control flow takes the true direction and \MAddOne is executing,
\PrimeProbe will evict
line \circled{2}.
As the control flow returns from \MAddOne and is about
to call \MDoubleOne, the program needs to fetch line \circled{2},
creating one  access in the midpoint of the iteration.
Then, while \MDoubleOne is executing, \PrimeProbe evicts
cache line \circled{2} again,
triggering a code fetch access when
returning from \MDoubleOne and executing the \texttt{if} statement.

Therefore, we   observe two accesses to line \circled{2} per iteration
if the bit value is 1,
and one access to line \circled{2}
if the bit value is 0.
It should be noted that, although line \circled{2}
slightly overlaps with the beginning of the \texttt{else} block,
we will \emph{not} observe an extra access if the bit value is 0.
This is because the overlapped region is executed immediately
after the \texttt{if} statement,
and the interval is too brief to be detected.

In practice, when we collect a trace of the memory accesses to the
target SF set
to which cache line \circled{2} maps, we also want to
collect the ground truth of nonce bit $k$ and iteration boundaries for validation purpose.
This requires some slight instrumentation of the binary,
a practice also seen in prior work~\cite{tlbleed,binoculars}.
The instrumentation is purely for validation purpose and
it is not necessary for the attack. However, due to
the instrumentation, the layout of the code changes, and it is easier to
monitor the cache line corresponding to the \texttt{else} direction.
The reasoning is similar to the explanation for line \circled{2}, but
we now observe the additional memory access at the midpoint of
an iteration when the bit value is 0, not 1.

\begin{figure}[t]
    \centering
    \includegraphics[width=\linewidth]{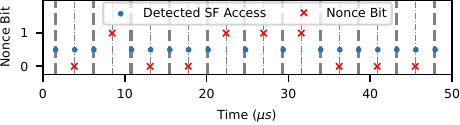}
    \caption{A snippet of memory accesses to the target SF set collected on \CR.
    Dots are  detected accesses,
and crosses are the nonce bit $k$ values (1 or 0).}
    \Description{A snippet of memory accesses
                 to the target SF set collected on \CR.}
    \label{fig:mont-trace}
    \vspace{-3mm}
\end{figure}

We collect the trace of memory accesses to the target SF set
(using the techniques of Section~\ref{sec:psd}),
the ground truth of nonce bit $k$,
and iteration boundaries on \CR,
while the victim code is executing.
Figure~\ref{fig:mont-trace} shows a short snippet of the trace that
happens to contain no noisy accesses made by other tenants.
In the figure,
thick dashed vertical lines represent the ground truth for iteration boundaries,
and thin dashed vertical lines represent halves of iterations.
Dots are  detected accesses,
and crosses are the nonce bit $k$ values (1 or 0).
Iterations where bit $k$ value is 0
have two accesses.  From the trace, we can easily read the nonce bits.

It takes only about \MontAvgIterationDuration on \CR to execute
one iteration of the Montgomery ladder loop that we target.
Thus, when the nonce bits have a sequence of continuous zeros,
the attacker needs to detect a sequence of accesses
that are \MontAvgHalfIterationDuration apart.
As shown in Table~\ref{tab:covert-comp}, the prime pattern of
\PrimeScope's~\cite{primescope} \PSFlush
takes on average   \CovertResPSFlushPrimeLat cycles to complete, while
the \PSAlt pattern has a low detection rate (Figure~\ref{fig:covert}).
As a result, the
\PrimeScope versions either frequently miss  memory accesses
or report  an access as occurring at a time different from when the actual access occurs.
In contrast,
our {\em Parallel Probing} strategy takes on average only \CovertResParallelPrimeLat
cycles  to execute (Table~\ref{tab:covert-comp}) and thus accurately detects the memory accesses
in ECDSA.

\vspace{-3mm}
\subsection{Finding the Target Cache Set with PSD}
\label{sec:extraction:finding}

We apply our PSD method to identify the victim's target SF set on \CR.
To obtain the ground truth,
we run the victim and attacker programs in the same container.
The attacker \texttt{mmap}s the victim program so that
the attacker can access the target line.
Then, when the attacker identifies an eviction set
that might correspond to the target SF set,
the attacker can validate it by checking
whether the eviction set indeed evicts the target line.

\parans{Scanning strategy}
Since the attacker knows the VA of the target cache line of the ECDSA victim,
they only need to construct eviction sets for SF sets at
the page offset of the target line and scan only those sets---i.e.,
it is the \PageOffset scenario.
To approximate the \WholeSF scenario,
we also measure the effectiveness of our approach by
scanning cache sets at every page offset in a \emph{random} order.

The ECDSA victim program
 spends only about   \MontExecutionRatio of its execution time running the vulnerable code.
Therefore,
there is a high chance that the attacker cannot detect the target set,
as they may collect the traces  while the victim is not executing the vulnerable
code---a problem known as de-synchronization.
Hence, the attacker repeatedly scans all possible sets until
detecting the target set or timeout.
We set the  timeouts for \PageOffset and \WholeSF to $\SI{60}{\second}$ and $\SI{900}{\second}$, respectively.
Time spent on eviction set construction is not counted towards the timeout.
\ifx\extended\undefined
Implementation details are presented
in the extended version of this paper~\cite{LLCFeasibleExt}.
\fi

\ifdefined\extended
\para{Scanner implementation}
To automatically detect the target set with the PSD method,
we train a supporting-vector machine (SVM)
to predict if a trace is from the target set based on the trace's PSD.
The SVM model
uses a polynomial kernel and
is trained using scikit-learn~\cite{scikit-learn}.
To train the model,
we collect \MontConfigNumPosPSDTrace traces from monitoring the target set
and \MontConfigNumNegPSDTrace traces from non-target sets from different \CR hosts.
We randomly withhold $30\%$ of the traces as the validation set and
use the remaining traces to train the model.
Our model has a \MontConfigPSDFNRate false-negative rate
and a \MontConfigPSDFPRate false-positive rate on the validation set.

During an attack, the attacker program first builds eviction sets
for the SF sets at the target page offset or in the whole system.
It then collects a $\SI{500}{\micro\second}$ access trace from each SF set.
In our implementation,
the attacker program running on \CR needs to stream access traces
back to a local machine to be processed by a Python program.
To reduce the data transmission overhead,
the attacker program performs a preliminary filtering and
only sends back traces containing 50-400 accesses.
This filtering range is empirically determined based on the victim's behavior
and the access trace length.
For every filtered trace,
the local Python program computes its PSD and uses
the SVM model to predict if it is from the target set.

Note that our PSD method may falsely identify a non-target set as a target set.
This can occur when scanning an SF set corresponding to the data
or instruction accesses performed by \MAdd or \MDouble,
as those accesses may occur at a frequency
that is similar to that of the target cache line.
To filter out these false positive results,
we attempt to extract the nonce bits from a positive access trace
using the method detailed in Section~\ref{sec:extraction:end2end}.
If we fail to extract enough nonce bits,
or if the extracted ``nonce bits'' are heavily biased towards $0$ or $1$,
we disregard this access trace and continue searching.
As we find that the risk of false positives is low for \PageOffset,
we only apply this technique to \WholeSF.
\fi

\parans{Evaluation setup}
We conduct  this experiment on \CR at different times of day,
totaling \MontConfigPageOffsetNumMeasure measurements for \PageOffset
and \MontConfigWholeSetNumMeasure measurements for \WholeSF.
For \WholeSF, we deem the scan successful if it manages to locate
the cache set accessed by the either side of the branch,
as accesses made by either side can disclose  the nonce $k$.

\parans{Results}
Table~\ref{tab:scanning} lists the key metrics of
finding the target cache set using the PSD method.
Given our timeout configurations,
\MontResPageOffsetSuccRate and \MontResWholeSetSuccRate of the scanning attempts
find the target set under \PageOffset and \WholeSF, respectively.
The lower success rate
under \WholeSF is mainly because
we can only scan  each SF set fewer times within the timeout period, %
leading to more failures due to the de-synchronization problem.
Averaged among successful scans,
it takes \MontResPageOffsetAvgDuration and \MontResWholeSetAvgDuration
to find the target set under \PageOffset and \WholeSF, respectively.
Finally, we scan from \SI[per-mode=symbol]{\MontResWholeSetAvgScanRateNoSpace}{\sets\per\second}
to \SI[per-mode=symbol]{\MontResPageOffsetAvgScanRateNoSpace}{\sets\per\second}.
The scanning speed can be improved
by using multiple threads to scan cache sets in parallel.

\begin{table}[t]
    \centering
    \caption{Performance of identifying the target cache set.}
    \label{tab:scanning}
    \small
    \setlength{\tabcolsep}{4pt}
    \begin{tabular}{|l|c|c|}
    \hline
    \textbf{Metric} & \textbf{\PageOffset} & \textbf{\WholeSF} \\ \hline
    Success Rate       & \MontResPageOffsetSuccRate & \MontResWholeSetSuccRate \\ \hline
    Average Success Time  & \MontResPageOffsetAvgDuration & \MontResWholeSetAvgDuration \\ \hline
    Std. Deviation of Success Time  & \MontResPageOffsetStdDuration & \MontResWholeSetStdDuration \\ \hline
    95\% Percentile Success Time  & \MontResPageOffsetNineFiveDuration & \MontResWholeSetNineFiveDuration \\ \hline
    Average Scan Rate   & \SI[per-mode=symbol]{\MontResPageOffsetAvgScanRateNoSpace}{\sets\per\second}
                     & \SI[per-mode=symbol]{\MontResWholeSetAvgScanRateNoSpace}{\sets\per\second}                  \\ \hline
    \end{tabular}
    \vspace{-4mm}
\end{table}

\subsection{End-to-End Nonce Extraction}
\label{sec:extraction:end2end}

Putting all the pieces together,
we demonstrate end-to-end, \emph{cross-tenant} nonce $k$ extractions on \CR.
In this demonstration,
the attacker first successfully co-locates their attack container
with the victim container~\cite{everywhereAllAtOnce}.
Then, the attacker builds the eviction sets
and finds the target set
using the PSD method,
while sending requests to trigger victim executions.
Once the target set is identified,
the attacker triggers the victim execution 10 more times
to steal the different nonces used in each execution.

To process the memory access trace,
we train a random forest classifier~\cite{pal2005random,scikit-learn}
to predict if a detected memory access corresponds to an iteration boundary.
To filter out false-positive boundary predictions,
we consider only boundary pairs that are
$8k$ to $12k$ cycles apart,
as this is the duration variation that
we expect from a single iteration on these hosts.
From each pair of predicted neighboring boundaries,
we recover the nonce bit in the iteration by checking
if there is an extra access in the middle of the iteration.

We attempt end-to-end nonce $k$ extractions
under the \PageOffset scenario on
\MontNumPairs pairs of co-located containers on \CR.
We identify a potential target set\tnote{
We have no oracle information to verify whether it is the target set under
this cross-container setting.
} and observe a signal in \MontNumPairsSignal of them.
Within the \MontNumTraces traces collected from these \MontNumPairsSignal victims,
we  extract an average of \MontAvgNonceRecover (or
a median value of \MontMedianNonceRecover) of the nonce bits.
Among these recovered bits,
our average bit error rate is \MontAvgBitErrorRate.
The full attack,
which includes constructing eviction sets,
identifying the target SF set,
and collecting $10$ traces,
takes an average of \MontAvgEndToEndDuration.

\vspace{-2mm}
\section{Related Work}
\label{sec:related}

\noindent{\textbf{Side-channel attacks in cloud.}}
Ristenpart~et~al.~\cite{getOffMyCloud} examined
the placement of virtual machines on physical hosts within AWS
and developed techniques to achieve co-location.
Zhang~et~al.~\cite{zhang2014crossTenant} employed \FlushReload
for a cross-tenant attack on a Platform-as-a-Service (PaaS) cloud.
However, %
\FlushReload is no longer feasible in modern clouds~\cite{remoteDedup,awsNitro}.
{\.{I}}nci~et~al.~\cite{inci2015seriously} in 2015 conducted a \PrimeProbe attack
on AWS EC2 to extract RSA keys,
using a reverse-engineered LLC slice hash function
and huge pages to build eviction sets.
Their attack is long running, relies on huge pages, and targets an
inclusive LLC---all of which are incompatible with modern cloud environments.

\noindent{\textbf{Mitigations to cache-based side-channel attacks.}}
Defenses can be broadly categorized into two types.
The first type, partition-based solutions~\cite{Catalyst,
secdcp,dawg,HybCache,bespoke,composableCachelets,chunkedCache,untangle},
blocks   attacks by partitioning the cache between different tenants.
However, this approach often requires complex hardware design and
results in high execution overhead.
The second type, randomization-based defenses~\cite{CeaserS,ceaser,ScatterCache,
saileshwar2021mirage,liu2014randomFill,Tan2020PhantomCacheOC,RPcache,Newcache,song2021randomized},
focuses on obfuscating the victim's cache usage.  %
While this method offers high performance,
it fails to provide comprehensive security guarantees.

\noindent{\textbf{Eviction set construction.}}
Algorithms for constructing eviction sets have received significant attention~\cite{LLCPractical,
EVSetTheoryPractice,song2019dynamically,RandomCacheAnalysis,xue2023ctpp,guo2022leaky}.
However, most approaches are developed and evaluated in a quiescent local environment.
Besides the group testing~\cite{EVSetTheoryPractice,CeaserS} and \PrimeScope~\cite{primescope}
algorithms discussed in Section~\ref{sec:back:evset:algo},
Prime+Prune+Probe (PPP)~\cite{RandomCacheAnalysis} exploits the LRU replacement policy
to  defeat  randomized caches by minimizing memory accesses.
CTPP~\cite{xue2023ctpp}, which is concurrent to our work,
builds on PPP by integrating it with \PrimeScope.
Based on the evaluation in CTPP~\cite{xue2023ctpp},
the success rates of both PPP and CTPP fall to almost zero
when a  single  memory-intensive SPEC 2006 benchmark~\cite{spec2006},
such as \texttt{mcf}, runs in the background.
Using the average LLC access rate as a metric,
the cache activity caused by \texttt{mcf} is only about $10\%$ of what we observed on \CR.
Lastly, Guo~et~al.~\cite{guo2022leaky} exploited a non-temporal prefetch instruction
to accelerate eviction set construction on Intel inclusive LLCs,
but found this technique inapplicable to Intel non-inclusive LLCs.

\noindent{\textbf{\PrimeProbe techniques.}}
Prior arts~\cite{libtea,armageddon} also used
parallel probing in their \PrimeProbe implementations~\cite{libteaPrimeProbeSrc,ARMageddonProbeSrc}.
However, to our knowledge,
we are the first to study the parallel probing strategy
to strike a good balance between probe and prime latency.
Oren~et~al.~\cite{oren2015spy} processed memory access traces in the frequency domain to fingerprint websites.

\vspace{-3mm}
\section{Conclusion}
\label{sec:conclusion}

In this paper, we presented an end-to-end, cross-tenant LLC \PrimeProbe attack
on a vulnerable ECDSA implementation in the public FaaS
Google Cloud Run environment.
We showed that  state-of-the-art eviction set
construction algorithms are ineffective on \CR.
We then introduced L2-driven candidate address filtering and
a binary search-based algorithm for address pruning to
speed-up eviction set construction.
Subsequently,
we introduced parallel probing to monitor  victim memory accesses
with high time resolution.
Finally, we leveraged power spectral density  
to identify the victim's target cache set in the frequency domain. 
Overall,
we extract a median value of
\MontMedianNonceRecover of the secret ECDSA nonce bits
from a victim container in \MontAvgEndToEndDuration on average.

\parans{Ethical considerations}
We limited our attempts to %
exfiltrate information from only victims %
under our control.
We monitored just one SF set of the host at a time,
thus minimizing potential performance interference with other tenants.

\vspace{-3mm}

\begin{acks}
We thank our shepherd, Yuval Yarom.
This work was funded in part by an Intel RARE gift;
by ACE, one of the 7 centers in JUMP 2.0,
an SRC program sponsored by DARPA;
and by NSF grants 1942888, 1954521, 1956007, 2154183, and 2107470.
\end{acks}

\balance
\bibliographystyle{ACM-Reference-Format}
\bibliography{refs}

\ifdef{\extended}{
  \appendix
  \section{Our Optimizations to Existing Eviction Set Construction Algorithms}
\label{app:algo-opt}

\para{Group testing}
In the baseline group testing algorithm~\cite{EVSetTheoryPractice},
the candidate set is divided into $G$ similarly-sized groups.
The algorithm temporarily withholds a group and tests
whether the remaining addresses can still evict a target address $\Taddr$
(Section~\ref{sec:back:evset:algo}).
Once such a removable group is found,
the algorithm prunes the identified group,
stops searching the remaining groups,
and re-partitions the remaining candidates into $G$ \emph{smaller} groups.
This approach is called early termination.
Prior work~\cite{song2019dynamically} suggests that this approach improves performance.
However, we find that an alternative approach that
continues to search the remaining groups without early termination
offers better performance and higher success rate on \SKX.
We believe that not performing early termination
helps the algorithm to prune away addresses more quickly
by pruning groups of larger sizes.
We call the alternative approach without early termination \GTOpt. %

We also implement a group testing variant suggested by Song~et~al.~\cite{song2019dynamically}.
This variant randomly withholds $n/W$ candidate addresses and
tests whether they are removable,
where $n$ is the remaining number of addresses in the candidate set.
Based on our evaluation,
this variant has similar performance and success rate as \GTOpt in
both local and \CR environments.
Therefore, we omit the discussion of this variant in the paper.

\para{\PrimeScope}
In the baseline \PrimeScope~\cite{primescope},
when the algorithm finds a congruent address,
the address is removed from the candidate set.
This depletes the congruent addresses that are near the head of the candidate list,
causing the algorithm to search deep into the list.
In \PSOpt, we gradually move candidate addresses from the back of the list
to the a near front position after we identify a congruent address.
This optimization ``recharges'' the front of the list
with more congruent addresses and
reduces the number of candidates being checked.

}

\end{document}